\documentclass[11pt]{article}

\usepackage{amsfonts,psfrag,epsfig}
\usepackage{amsfonts,epsfig}
\usepackage{color}
\usepackage{amsmath,amssymb}
\usepackage[amsmath,thmmarks]{ntheorem}
\usepackage[margin = 1.1in]{geometry}

\newtheorem{theorem}{Theorem}
\newtheorem{lemma}[theorem]{Lemma}
\newtheorem{corollary}[theorem]{Corollary}
\newtheorem{proposition}[theorem]{Proposition}

\newtheorem{definition}{Definition}

{ \theoremstyle{nonumberplain}\theoremsymbol{\ensuremath{\Box}}
\theorembodyfont{\normalfont}
\newtheorem{proof}{Proof.}

\theoremsymbol{\ensuremath{\Box}}
\theoremheaderfont{\normalfont\itshape}
\theorembodyfont{\normalfont}
\theoremstyle{empty}

}

\newcommand{\union}{\cup}

\renewcommand{\Pr}{{\mathbb P}}

\newcommand{\norm}[1]{\left\lVert #1 \right\rVert}

\newcommand{\paraG}[1]{\paragraph{#1.}}

\newcommand{\bind}{\bold{I}}

\newcommand{\expo}[1]{\exp\left( #1 \right)}

\newcommand{\bW}{\boldsymbol{W}}

\newcommand{\ba}{\boldsymbol{a}}
\newcommand{\bA}{\boldsymbol{A}}

\newcommand{\sig}{\boldsymbol{\sigma}}

\newcommand{\brho}{\boldsymbol{\rho}}
\newcommand{\bmu}{\boldsymbol{\mu}}
\newcommand{\bnu}{\boldsymbol{\nu}}
\newcommand{\bpi}{\boldsymbol{\pi}}

\newcommand{\bsigma}{\boldsymbol{\sigma}}
\newcommand{\bdelta}{\boldsymbol{\delta}}
\newcommand{\bdel}{\bdelta}
\newcommand{\rhO}{\boldsymbol{\rho}}
\newcommand{\lamb}{\boldsymbol{\lambda}}
\newcommand{\Lamb}{\boldsymbol{\Lambda}}

\newcommand{\sx}{{\sf x}}

\newcommand{\cN}{\mathcal{N}}

\newcommand{\cT}{\mathcal{T}}

\newcommand{\qpi}{\boldsymbol{\xi}}
\newcommand{\mix}{\text{mix}}

\newcommand{\beq}{\begin{eqnarray}}
\newcommand{\eeq}{\end{eqnarray}}
\newcommand{\beqn}{\begin{equation}}
\newcommand{\eeqn}{\end{equation}}
\newcommand{\R}{\mathbb{R}}
\newcommand{\N}{\mathbb{N}}
\newcommand{\Z}{\mathbb{Z}}

\newcommand{\Rp}{\mathbb{R}_+}
\newcommand{\Zp}{\mathbb{Z}_+}

\newcommand{\bzero}{\mathbf{0}}
\newcommand{\bone}{\mathbf{1}}

\newcommand{\mc}{\mathcal}
\newcommand{\mb}{\mathbf}

\newcommand{\cM}{\mc{M}}
\newcommand{\sX}{{\sf X}}

\newcommand{\cW}{\mc{W}}

\newcommand{\cE}{\mc{E}}
\newcommand{\cG}{\mc{G}}

\newcommand{\cI}{\mc{I}}

\newcommand{\bB}{\mathbf{B}}
\newcommand{\bu}{\mb{u}}
\newcommand{\bv}{\mb{v}}

\newcommand{\bx}{\mb{x}}

\newcommand{\by}{\mb{y}}

\newcommand{\bQ}{\mathbf{Q}}

\newcommand{\E}{\mathbb{E}}

\newcommand{\y}{\boldsymbol{y}}

\DeclareMathOperator{\Conv}{\textsf{Conv}}

\begin{document}

\title{Efficient Distributed Medium Access
\footnote{Work of D.\ Shah and J.\ Shin supported in parts by the
AFOSR complex networks grant and DARPA ITMANET grant. Research of
P.\ Tetali supported in part by NSF grant CCF-0910584.} }


\author{Devavrat Shah\footnote{Laboratory for
Information and Decision Systems and Department of Electrical
Engineering and Computer Science at Massachusetts Institute of
Technology. Email: devavrat@mit.edu. } \and Jinwoo
Shin\footnote{Algorithms and Randomness Center at Georgia Institute
of Technology. Email: jshin72@cc.gatech.edu.} \and Prasad
Tetali\footnote{School of Mathematics and School of Computer Science
at Georgia Institute of Technology. Email: tetali@math.gatech.edu.}}

\maketitle

\thispagestyle{empty}

\begin{abstract}

Consider a wireless network of $n$ nodes represented by a graph
$G=(V, E)$ where
an edge $(i,j) \in E$ models the fact that transmissions of $i$ and
$j$ {\em interfere} with each other, i.e.\ simultaneous
transmissions of $i$ and $j$ become unsuccessful. Hence it is
required that at each time instance a set of non-interfering nodes
(corresponding to an independent set in $G$) access the wireless
medium. To utilize wireless resources efficiently, it is required to
arbitrate the access of medium among interfering nodes properly.
Moreover, to be of practical use, such a mechanism is required to be
totally distributed as well as simple.

As the main result of this paper, we provide such a medium access
algorithm. It is randomized, totally distributed and simple: each
node attempts to access medium at each time with probability that is
a function of its local information. We establish efficiency of the
algorithm by showing that the corresponding network Markov chain is
positive recurrent as long as the demand imposed on the network can
be supported by the wireless network (using {\em any} algorithm). In
that sense, the proposed algorithm is optimal in terms of utilizing
wireless resources. The algorithm is oblivious to the network graph
structure, in contrast with the so-called {\em polynomial back-off}
algorithm by Hastad-Leighton-Rogoff (STOC '87, SICOMP '96) that is
established to be optimal for the complete graph and bipartite
graphs (by Goldberg-MacKenzie (SODA '96, JCSS '99)).

\end{abstract}



\newpage
\setcounter{page}{1}

\section{Introduction}\label{sec:model}

We consider a {\em single-hop} wireless network of $n$ nodes or
queues  represented by $V = \{1,\dots, n\}$. Time is discrete
indexed by $\tau \in \{0,1,\dots\}$. Unit-size packets arrive at
queue $i$ as per an exogenous process. Let $A_i(\tau)$ denote the
number of packets arriving at queue $i$ in the time slot
$[\tau,\tau+1)$. For simplicity, we shall assume $A_i(\cdot)$ as an
independent Bernoulli process with rate $\lambda_i$, i.e.\
$\lambda_i = \Pr(A_i(\tau)= 1)$ and $A_i(\tau) \in \{0,1\}$ for all
$i, \tau \geq 0$.\footnote{The result in this paper extends easily
even for (non-Bernoulli) adversarial arrival processes satisfying
$\sum_{\tau=s}^{t-1} A_i(\tau)\leq \lambda_i(t-s)+w$, with (fixed)
$w<\infty$, for all $0\leq s<t$.
} Let $Q_i(\tau) \in \N$ be the
number of packets in queue $i$ at the beginning of time slot
$[\tau,\tau+1)$.

The work from queues is served at the unit rate subject to {\em
interference} constraints. Specifically, let $G = (V,E)$ denote the
inference graph : $(i,j) \in E$ implies that queues $i$ and $j$ can
not transmit simultaneously since their transmissions {\em
interfere} with each other. Formally, let $\sigma_i(\tau) \in
\{0,1\}$ denote whether the queue $i$ is (successfully) transmitting
at time $\tau$, and let $\sig(\tau) = [\sigma_i(\tau)]$. Then,
$$\sig(\tau) \in \cI(G) \stackrel{\Delta}{=}
\{ \rhO = [\rho_i] \in \{0,1\}^n :
   \rho_i + \rho_j \le 1\text{ for all }(i,j) \in E \},\quad\mbox{for}~\tau\geq0,$$
i.e.\ $\cI(G)$ is the set of independent sets of $G$.  The resulting
queueing dynamics can be summarized as
$$ Q_i(\tau+1) = Q_i(\tau) - \sigma_i(\tau) \bind_{\{Q_i(\tau) > 0\}}
+ A_i(\tau), ~~\text{for}~\tau\geq0, ~1\leq i\leq n. $$ Here
$\bind_{\{x\}} = 1$ if $x =$`true' and $0$ if $x=$`false'.

Now an algorithm, which we shall call {\em medium access}, is
required to choose $\bsigma(\tau) \in \cI(G)$ at the beginning of
each time $\tau$.  A good medium access algorithm should choose
$\bsigma(\tau)$ so as to utilize the wireless medium as efficiently
as possible. Putting it another way, it should be able to keep
queues finite for as large a set of arrival rates
$\lamb=[\lambda_i]$ as possible. Towards this, define the capacity
region
\begin{align*}
 \Lamb & =\Big\{ \y \in \Rp^n : \y<\sum_{\sig \in \cI(G)} \alpha_{\sig} \sig\text{ with }\alpha_{\sig}\geq 0,
\sum_{\sig \in \cI(G)} \alpha_{\sig} \leq 1 \Big\}.
\end{align*}
Since $\sig(\tau) \in \cI(G)$, the effective `service' rate induced
by any algorithm over time is essentially in the closure of $\Lamb$.
Therefore, a medium access algorithm can be considered optimal, if
it can keep queues {\em finite, for any} $\lamb \in \Lamb$.
Formally, if the state of the queueing system including the
algorithm's decisions and queue-sizes can be described as a Markov
chain, then the existence of a stationary distribution for this
Markov chain and its ergodicity effectively implies that the queues
remain finite. A sufficient condition for this is {\em aperiodicity}
and {\em positive recurrence} of the corresponding Markov chain.
This motivates the following definition.
\begin{definition}[Optimal] {\em A medium access algorithm is
called {\em optimal} if for any $\lamb \in \Lamb$ the (appropriately defined)
underlying network Markov chain is \emph{positive recurrent} and {\em aperiodic}}.
\end{definition}
To be of practical use, medium access algorithm ought to be simple
and totally distributed, i.e.\ should use only local information
like queue-size, and past collision history. In such an algorithm,
each node makes the decision to transmit or not on its own, at the
beginning of each time slot. At the end of the time slot, it learns
whether attempted transmission was successful or not (due to a
simultaneous attempt of transmission by a neighbor). If a node does
not transmit,  then it learns whether any of its neighbors attempted
transmission. And, ideally such an algorithm should be optimal.

\subsection{Prior Work}

Design of an efficient and distributed medium access algorithm has
been of interest since the ALOHA algorithm for the radio network
\cite{Abramson} and Local Area Networks \cite{MB76} in the 1970s.
Subsequently a variety of the so-called back-off algorithms or
protocols have been extensively studied. Various negative and
positive properties of back-off protocols were established in
various works \cite{Kel85, MH85, KM87, TL87, Ald87, MacP89}.

Specifically, Hastad, Leighton and Rogoff \cite{HLR96} studied a
medium access algorithm in which each node or queue attempts
transmission at each time with probability that is inversely
proportional to a polynomial function of the number of successive
failures in the most recent past. They established it to be optimal
when the interference graph $G$ is complete, or equivalently all
nodes are competing for one resource (as in the classical
Ethernet/LAN). The optimality of this polynomial back-off algorithm
was further established for $G$ when it is induced by matching
constraints in a bipartite graph by Goldberg and MacKenzie
\cite{Goldberg1999232}. However, the optimality of polynomial
back-off or any other totally distributed medium access algorithm
remained open for general graphs. The interested reader may find a
good summary of results until 2001, on medium access  on a webpage
maintained by Goldberg \cite{LAG}.

In the past year or so, significant progress has been made towards
this open question. Specifically, Rajagopalan, Shah and Shin (RSS)
\cite{RSS09, SS09} and Jiang and Walrand (JW) \cite{JW08} proposed
two different medium access algorithms that operate in continuous
time. The RSS algorithm is optimal but requires a bit of information
exchange between each pair of neighboring nodes per unit time. The
JW algorithm was established to have a weaker form of optimality,
called `rate stability', by Jiang, Shah, Shin and Walrand
\cite{JSSW10}. In summary, both algorithms stop short of being
totally distributed and optimal. Further, both of them operate in
continuous time and thus effectively avoiding the issue of loss in
performance due to contention present in discrete time considered in
this paper.

\subsection{Our Contribution}\label{sec:contrib}

The main result of this paper is a totally distributed medium access
algorithm that is optimal for any interference graph $G$. It
resolves an important question in distributed computation that has
been of great practical interest. The proposed medium access
algorithm builds on the RSS algorithm and overcomes its two key
limitations by adapting it to the discrete time and removing the
need for any information exchange between neighboring nodes.
In what follows, we explain in detail how we overcome such
limitations.

In the proposed medium access algorithm, each node attempts
transmission in each time slot based on: (a) whether it managed to
successfully transmit in the previous time slot, or whether any of
its neighbors attempted to transmit in the previous time slot; (b)
local queue-size and estimation of the ``weight" of the neighbors.
Given this information, each node in the beginning of each time slot
attempts transmission with probability depending upon (a) and (b).
Specifically, if the node was successful in the previous time, it
does not transmit in this time with probability that is inversely
proportional to its own weight that depends on (b). Else if no other
neighbor attempted transmission in the previous time then a node
attempts transmission with probability $\frac12$. Otherwise, with
probability $1$, a node does not transmit.

In such an algorithm, the only seeming non-local information is the
estimation of the neighbors' weight in (b). An important
contribution of this work is the design of a non-trivial learning
mechanism, based only on information of type (a), that estimates the
neighbors' weight without any explicit information exchange. We note
that, in contrast, the RSS algorithm had required explicit
information exchange for estimating the neighbors' weight.

To establish optimality of the proposed algorithm, we show  that,
in essence, the value of $\sum_i \sigma_i(\tau) \log\log Q_i(\tau)$
is {\em close} to $\max_{\brho\in \cI(G)} \sum_i \rho_i \log\log
Q_i(\tau), $ on average for all large enough $\tau$. That is,
effectively the distributed medium access chooses $\sigma(\tau)$
that is (close to) maximum weight independent set of $G$ when node
weights are equal to $\log \log $ of the queue-sizes. Such a
property is known (cf.\ \cite{SW06, stolyar}) to imply that $\sum_i
F(Q_i(\tau))$ (where $F(x) = \int_0^x \log\log y~ dy$) is a
potential (or Lyapunov, energy) function for the system state so
that the function is expected to decrease by at least a fixed amount
as long as $\lamb \in \Lamb$. This subsequently establishes that the
network as a Markov chain is positive recurrent (implying  the
optimality of the algorithm).

We establish the near optimality of $\sum_i \sigma_i(\tau) \log\log
Q_i(\tau)$ under the medium access algorithm in two steps. To begin
with, we observe that the evolution of $\sig(\tau)$ under the
algorithm is a Markov chain on the space of independent sets
$\cI(G)$ with time-varying transition probabilities. For this Markov
chain, at any particular time $\tau$, let $\bpi(\tau)$ be the
stationary distribution at time $\tau$ (given transition
probabilities at time $\tau$).

In the first step, we study this (time-varying, stationary)
distribution $\bpi(\tau)$ and show that it is approximately
`product-form'. To obtain such an approximate characterization, we
show that the transition probabilities of the Markov chain are well
approximated by those of a reversible Markov chain which has a
product-form stationary distribution. A novel comparison relation
between stationary distributions of two Markov chains in terms of
the relation between their transition probabilities leads to the
approximate product-form characterization of $\bpi(\tau)$. We note
that  the RSS algorithm (and similarly, the JW algorithm) had used
the continuous time setting to make sure that the corresponding
Markov chain was reversible and hence had a product-form
distribution to start with; such reversibility is lost in general in
the discrete time setting of this paper.
Using this approximate product-form characterization of $\bpi(\tau)$
in addition to the Gibbs' maximal principle (cf.\ \cite{GBook}), we
prove that $\bpi(\tau)$ has the desired property; namely, that
$\sum_i \sigma_i \log\log Q_i(\tau)$ is close to $\max_{\brho\in
\cI(G)} \sum_i \rho_i \log\log Q_i(\tau)$ if $\sig=[\sigma_i]$ is
given by  the distribution $\bpi(\tau)$. We call this  the
maximum-weight property at stationarity.

In the second step,  we show that the  Markov chain, despite it
being time-varying, is always near stationarity for large enough
$\tau$ by carefully estimating the effective mixing time of the
time-varying Markov chain. In other words, the distribution of
$\sig(\tau)$ is close to $\bpi(\tau)$ for large enough $\tau$.
Therefore, the maximum-weight property at stationarity (established
in the first step) implies that $\sum_i \sigma_i(\tau) \log\log
Q_i(\tau)$ is close to $\max_{\brho\in \cI(G)} \sum_i \rho_i
\log\log Q_i(\tau)$. To guarantee the near stationarity property as
a consequence of such a mixing analysis, it is required that a
design of `weight' maintained by each node in the medium access
algorithm utilizes the neighbor's weight information. As mentioned
earlier, we resolve this by developing a learning mechanism that
estimates the neighbor's weight based on the information whether it
transmitted or not thus far. The success in this second step is
primarily due to our novel design of the learning mechanism
incorporated well with the mixing time analysis of the time-varying
Markov chain.


\subsection{Organization}

Remainder of the paper is organized as follows. Section
\ref{sec:algo} presents formally the medium access algorithm and a
statement of the main result. Section \ref{sec:pfsum} summaries our
proof strategy for the main result at a high level. Section
\ref{sec:prelim} presents necessary preliminaries that are used to
establish the main result in Section \ref{sec:proof}. Section
\ref{sec:proof.a} and \ref{sec:proof.b} are for providing the
detailed proof of the main lemma in Section
\ref{sec:proof}. 

\section{Algorithm and Its Optimality}\label{sec:algo}

The medium access algorithm is randomized, distributed, simple and
runs in discrete time with time indexed by $\tau \geq 0$.  Each node
$i \in V$ maintains weight $W_i(\tau) \in \Rp$ over $\tau \geq 0$.
In the beginning of each time slot $\tau \geq 0$, each node $i \in
V$ decides to attempt transmission or not as follows:

\medskip
\begin{itemize}

\item[1.] If the transmission of node $i$ was successful at
$\tau-1$, then
\begin{itemize}
\item[$\circ$] it attempts to transmit with probability $1-\frac{1}{W_i(\tau)}$.
\end{itemize}

\item[2.] Else if no neighbor of $i$ attempted transmission at
$\tau-1$, then
\begin{itemize}
\item[$\circ$] it attempts to transmit with probability $\frac{1}{2}$.
\end{itemize}

\item[3.] Otherwise, it does not attempt to transmit with probability $1$.
\end{itemize}

\medskip

Now we describe how each node $i$ maintains weight $W_i(\tau)$:
\begin{equation}
W_i(\tau) ~=~\max\left\{\log Q_i(\tau), \max_{j\in \mathcal{N}(i)} \exp\Big({\sqrt{\log g(A^i_j(\tau))}}\Big)\right\},
\label{eq:weight}
\end{equation}
where by $\log$  and $\log\log$ we mean $[\log]_+$ and
$[\log\log]_+$  respectively; $g: \Rp\to\Rp$ is defined as $g(x) =
\exp(\log\log^4 x)$; by $\log\log^4 x$ we mean $(\log\log x)^4$,
$\log$ represents $\log_e$; and $\cN(i) = \{j \in V: (i,j) \in E\}$
represents neighbors of node $i$. Note that $W_i(\tau) \geq 1$
for all $\tau$ by definition. In above, $A^i_j(\cdot)$ is a
`counter' maintained by node $i$ as a `long term' estimate of weight
$W_j(\cdot)$. This is maintained along with another `counter'
$B_j^i(\cdot)$ by node $i$ as a `short term' estimate of
$W_j(\cdot)$. Initially, $A_j^i(0) = B_j^i(0) = 0$ for all $j \in
\cN(i)$ and $i \in V$. For each $j \in \cN(i)$, $A_j^i(\cdot)$ and
$B_j^i(\cdot)$ are updated by node $i$ at $\tau$ as follows:

\medskip

\begin{itemize}

\item[1.] If $j\in \cN(i)$ attempted transmission at $\tau-1$, then
\begin{itemize}
\item[$\circ$] $A^i_j(\tau)= A^i_j(\tau-1)$ and $B^i_j(\tau)= B^i_j(\tau-1) +1$.
\end{itemize}
\item[2.] Else if $B^i_j(\tau-1)\geq 2$, then
\begin{itemize}
\item[$\circ$] $A^i_j(\tau) =
\begin{cases}
A^i_j(\tau-1)+1 &\mbox{if}~~ B^i_j(\tau-1)\geq g(A^i_j(\tau-1))\\
A^i_j(\tau-1)-1 &\mbox{otherwise}\\
\end{cases}$ and reset $B^i_j(\tau)= 0$.
\end{itemize}
\item[3.] Otherwise, $A^i_j(\tau) = A^i_j(\tau-1)$ and $B^i_j(\tau)= 0$.
\end{itemize}

\medskip

Observe that $B^i_j(\cdot)$ counts how long neighbor $j$ keeps
attempting transmission consecutively. When $j$'s transmissions are
successful, the random period of consecutive transmissions is
essentially distributed as per the geometric distribution with mean
$W_j(\cdot)$ due to the nature of our algorithm. Thus $B^i_j(\cdot)$
provides a short-term (or instantaneous) estimation of $W_j(\cdot)$.
To extract a robust estimation of $W_j(\cdot)$ from such short-term
estimates, the long-term estimation $A^i_j(\cdot)$ is maintained: it
changes by $\pm 1$ using $B_j^i(\cdot)$ at most once per unit time.
Specifically, as per the above updates, $g(A^i_j(\cdot))$ is acting
as an estimation of $W_j(\cdot)$. Now it is important to note that
the choice of $g$ (defined above) plays a crucial role in the
quality of estimate of $W_j(\cdot)$. The change in estimation
$g(A_j^i(\cdot))$, when $A_j^i(\cdot)$ is updated by $\pm 1$, is
roughly $g'(A_j^i(\cdot))$. Since $W_j(\cdot)$ is changing over
time, it is important to have $g'(\cdot)$ not too small. On the
other hand, if it is too large then it is too sensitive and could be
noisy just like $B_j^i(\cdot)$. A priori it is not clear if there
exists a choice of any function $g$ that allows for keeping
$A_j^i(\cdot)$ for a good enough estimation of $W_j^i(\cdot)$, which
subsequently leads to positive-recurrence of the network Markov
chain. Somewhat surprisingly (at least to us), we find that indeed
such a $g$ exists and is as defined above: $g(x) = \exp(\log\log^4
x)$. As per our proof technique, $g(x)=\exp(\log\log^\alpha x)$
works for any $\alpha > 2$; however we shall stick to the choice of
$\alpha = 4$ in the paper. Section \ref{sec:pfsum} provides the
reasons on why such a choice of function $g$ is necessary and
sufficient. Now we state the main result of this paper.

\begin{theorem}\label{thm:main}
The medium access algorithm as described above is optimal for any
interference graph.
\end{theorem}

\section{Proof of Theorem \ref{thm:main}: An Overview}\label{sec:pfsum}

This section provides an overview of the proof of Theorem
\ref{thm:main} to explain the key challenges involved in
establishing it as well as intuition behind the particular choice of
function $g$. The goal in this section is not to provide precise
arguments but only provide intuition so as to assist a reader in
understanding the structure of the proof. The complete proof with
all details is stated in Sections \ref{sec:proof}, \ref{sec:proof.a}
and \ref{sec:proof.b}.

Theorem \ref{thm:main} requires establishing positive recurrence of an
appropriate Markov chain that describes the evolution of the network state
under the medium access algorithm described. To that end, define
\[
X(\tau) = (\bQ(\tau), \sig(\tau),\ba(\tau), \bA(\tau), \bB(\tau))
\]
where $\bQ(\tau)$ represents vector of queue-sizes; $\ba(\tau) \in \{0,1\}^n$
denotes the vector of transmission attempts by nodes at time $\tau$;
$\sig(\tau) \in \cI(G)$ denotes the vector of resulting successful
transmissions in time $\tau$ (clearly, $\sig(\tau) \leq \ba(\tau)$); and
$\bA(\tau), ~\bB(\tau) \in \Zp^{2|E|}$ denote the vector of
long-term and short-term estimations maintained at nodes as explained
in Section \ref{sec:algo}. Then it follows that under medium access algorithm
$X(\cdot)$ is a Markov chain.

Now a generic method to establish positive-recurrence of a Markov
chain involves establishing that certain real-valued function over
the state-space of the Markov chain is Lyapunov or Potential
function for the Markov chain. Roughly speaking, this involves
establishing that on average the value of this function decreases
under the dynamics of the Markov chain if its value is high enough;
Theorem \ref{thm:PR} states the precise conditions that need to be
verified. With this eventual goal, we consider the following
function that maps state $\sx = (\bQ,\sig,\ba,\bA, \bB)$ to
non-negative real values as
\begin{align}\label{eq:def.lyp}
L(\sx) & = \sum_i F(Q_i) + \sum_{i; j \in \cN(i)} \Big((A_j^i)^2 + g^{(-1)}(B^i_j) \Big),
\end{align}
where $F(x) = \int_0^x \log\log y~dy$ with $\log \log y = [\log\log
y]_+$; the inverse function of $g(x) =  \exp(\log\log^4 x)$ is
$g^{(-1)}(x) = \exp(\exp(\log^{1/4} x))$. With abuse of notation, we shall use $L(\tau)$
to denote $L(X(\tau))$. Now
\begin{eqnarray*}
L(\tau)&=&L^{Q}(\tau)+L^{A,B}(\tau),
\end{eqnarray*}
where $L^{Q}(\tau) ~= \sum_i F(Q_i(\tau))   ~\mbox{and} ~
L^{A,B}(\tau)=\sum_{i; j \in \cN(i)} \left((A_j^i(\tau))^2 +
g^{(-1)}(B^i_j(\tau)) \right).$

The proof is devoted to establish the {\em negative-drift} property
of $L(\cdot)$, i.e.\ if $X(\tau) = \sx$ is such that $L(\tau)$ is
{\em large enough} (i.e.\ larger than some finite constant), then
value of $L(\cdot)$ decreases enough on average. This property is
established by considering two separate cases.
\begin{itemize}
\item[] {\em Case One.} When $L(\tau)$ is large due to the component $L^{A,B}(\tau)$
being very large.
\begin{itemize}
\item[$\circ$]  Formally, when $\max_{i, j} \big(g(A_j^i(\tau)),
B_j^i(\tau)\big) \geq \bW^3_{\max}(\tau)$, where $\bW_{\max}(\tau) =
\max_i W_i(\tau)$.
\end{itemize}
\item[] {\em Case Two.} When $L(\tau)$ is  large due to the component  $L^{Q}(\tau)$ being
very large.
\begin{itemize}
\item[$\circ$]  Formally, when $\max_{i, j} \big(g(A_j^i(\tau)),
B_j^i(\tau)\big) < \bW^3_{\max}(\tau)$, where $\bW_{\max}(\tau) =
\max_i W_i(\tau)$.
\end{itemize}
\end{itemize}

\paraG{Case One} In this case, there exists $i \in V$ and $j \in \cN(i)$ so
that $g(A^i_j(\tau))$ or $B^i_j(\tau)$ is larger than
$\bW^3_{\max}(\tau)$. Using the property of the estimation procedure
(which updates $A_j^i(\cdot)$), we show that the $L^{A,B}(\cdot)$
decreases on average by a large amount; it is large enough so that
it dominates the possible increase in any other components of
$L(\cdot)$. Such a strong property holds because as per the
algorithm, $g(A^i_j(\tau))$ and $B^i_j(\tau)$ continually try to
estimate $W_j(\tau)$ and hence if either of them is larger than
$\bW_{\max}^3(\tau)$, they ought to decrease by a large amount in a
short time. Indeed, to translate this property into sufficient
decrease of $L(\cdot)$, the careful choice of $L^{A,B}(\cdot)$ is
made. This case is dealt in detail in Section \ref{sec:proof.a}.


\paraG{Case Two} In this case, for each $i \in V$ and $j \in \cN(i)$,
$g(A^i_j(\tau))$ and $B^i_j(\tau)$ are smaller than
$\bW_{\max}^3(\tau)$. To establish the decrease in $L(\cdot)$, we
show that in this case $L^Q(\cdot)$ decreases by large enough
amount; large enough so that it dominates the possible increase in
$L^{A,B}(\cdot)$. This case crucially utilizes the property of the
medium access algorithm, the choice of the weights $W_i(\cdot)$ for
$i \in V$ and the form of function $g$. The precise details
explaining how these play roles in establishing this decrease in
$L^Q(\cdot)$ is explained in Section \ref{sec:proof.b}. Here, we
shall provide key ideas behind these somewhat involved arguments.

The property that $L^Q(\cdot)$ decreases by large enough amount
follows if we establish that the set of transmitting nodes
$\sig(\tau)$ is such that
\begin{align}\label{eq:ov.1}
\sum_i \sigma_i(\tau) \log\log Q_i(\tau) & \approx \max_{\rhO\in\mathcal I(G)} \sum_i \rho_i \log\log  Q_i(\tau).
\end{align}
See Lemma \ref{lem:qh2} for further details.  To establish
\eqref{eq:ov.1}, using the condition of the second case
$g(A_j^i(\tau)) < \bW^3_{\max}(\tau)$ for all $i\in V$ and $j \in
\cN(i)$, we essentially show that
\begin{align}
g(A_j^i(\tau)) & \approx W_j(\tau), ~~\mbox{for~all}~i \in V, ~j\in \cN(i), ~\mbox{and}\label{eq:ov.2a} \\
\sum_i \sigma_i(\tau) \log\log Q_i(\tau)  & \approx  \max_{\rhO\in\mathcal I(G)} \sum_i \rho_i \log W_i(\tau) \label{eq:ov.2b}
\end{align}
To see why \eqref{eq:ov.2a} and \eqref{eq:ov.2b} are sufficient to
imply \eqref{eq:ov.1}, note that
\begin{align*}
\big|\log W_i(\tau)-\log\log Q_i(\tau)\big| &~\leq~  \max_{j\in
\mathcal{N}(i)} {\sqrt{\log g(A^i_j(\tau))}}
 ~\approx~ \max_{j\in \mathcal{N}(i)} {\sqrt{\log W_j(\tau)}} \\
& ~\ll~ \max_{\rhO\in\mathcal I(G)} \sum_i \rho_i \log W_i(\tau),
\end{align*}
when $\bW_{\max}(\tau)$ (or $\bQ_{\max}(\tau)$) is very large. Therefore,
\begin{align*}
\max_{\rhO\in\mathcal I(G)} \sum_i \rho_i \log\log Q_i(\tau) & \approx~  \max_{\rhO\in\mathcal I(G)} \sum_i \rho_i \log W_i(\tau).
\end{align*}
In summary, to establish desired decrease in $L^Q(\cdot)$, it boils down to establishing
\eqref{eq:ov.2a} and \eqref{eq:ov.2b}.

\medskip
To establish \eqref{eq:ov.2a}, it is essential for $g(\cdot)$ to be {\em growing fast enough}
so that if $g(A_j^i(\tau))$ is very different (in this case, smaller) compared to $W_j(\tau)$,
then under the execution of the algorithm, it quickly converges (close) to $W_j(\cdot)$. For this,
it is important that $g(A_j^i(\cdot))$ should change at a faster rate compared to the
rate at which $W_j(\cdot)$ changes.  Towards that, note that if $A^i_j(\cdot)$ is
updated (by unit amount) then  $g(A^i_j(\cdot))$ roughly changes by amount $g^{\prime}(A^i_j(\tau))$,
which is at least
$$g^{\prime}(A^i_j(\tau))> g^{\prime}(g^{(-1)}(\bW_{\max}(\tau)^3)).$$
Here we have used the fact that $g^{\prime}$ is a decreasing function and
$g(A^i_j(\tau))$ is at most $\bW_{\max}(\tau)^3$. Using properties of
function $g$, we establish that (see Proposition \ref{clm:wbound})
$W_j(\tau)$ changes per unit time by at most
$$\frac{W_j(\tau)}{g^{(-1)}\left(\expo{\log^2
W_j(\tau)}\right)}.$$ For the purpose of developing an intuition
regarding the choice of $g$, consider $j \in \arg\max_i W_i(\tau)$,
i.e.\ $W_j(\tau) = \bW_{\max}(\tau)$. Then, such a $W_j(\tau)$
changes as
$$\frac{W_j(\tau)}{g^{(-1)}\left(\expo{\log^2
W_j(\tau)}\right)} =
\frac{\bW_{\max}(\tau)}{g^{(-1)}\left(\expo{\log^2
\bW_{\max}(\tau)}\right)}.$$ Therefore, to have $g$ such that the
change in $W_j(\cdot)$ is slower than that in $g(A_j^i(\cdot))$, we
must have
$$g^{\prime}(g^{(-1)}(\bW_{\max}(\tau)^3))~>~ \frac{\bW_{\max}(\tau)}{g^{(-1)}\left(\expo{\log^2
\bW_{\max}(\tau)}\right)}.$$ Our interest will be having properties holding when
$\bW_{\max}(\tau)$ (or $\bQ_{\max}(\tau)$) is large enough. This leads to the
condition that
\begin{align*}
\lim_{x\to\infty} g^{\prime}(g^{(-1)}(x^3))
\frac{g^{(-1)}\left(\expo{\log^2 x}\right)}{x} & >~ 1.
\label{eq:ov.3}
\end{align*}
It can be checked that the above condition is satisfied if $g(x)$
does not grow slower than $\exp(\log\log^\alpha x)$ for some
constant $\alpha
> 2$.\footnote{We say $g$ does not grow slower and
faster than $f$ if $\lim_{x\to\infty} \frac{g(x)}{f(x)}>0$ and
$\lim_{x\to\infty} \frac{g(x)}{f(x)}<\infty$, respectively.} That
is, we need $g$ to be growing roughly at least as fast as the choice
made in the description of our algorithm in Section \ref{sec:algo}.
Precise details on how this choice of $g$ guarantees $g(A_j^i(\tau))
\approx W_j(\tau)$ is given in Section \ref{sec:proof.b}.

\medskip

Next, discussion on how we establish \eqref{eq:ov.2b}, which will
require another condition on $g(\cdot)$ to be {\em growing slow
enough}, in contrast to the fast enough growing condition for
\eqref{eq:ov.2a}. Effectively, we need to establish that
$\bmu(\tau)$, the distribution of $\sig(\tau)$ under the algorithm,
is concentrated around the subset of schedules with high-weight,
i.e.\ roughly speaking the subset
\begin{align}\label{eq:ov.4}
\Big\{ \tilde{\rhO}=[\tilde{\rho}_i] \in \cI(G) :  \sum_i
\tilde{\rho}_i \log W_i(\tau) \approx \max_{\rhO\in\mathcal I(G)}
\sum_i \rho_i \log W_i(\tau)\Big\}.
\end{align}
To that end, consider the evolution of schedule $\sig(\tau)=[\sigma_i(\tau)]$
and weight $\bW(\tau)=[W_i(\tau)]$ under the algorithm. Now the
distribution of $\sig(\tau)$ depends on the schedule
$\sig(\tau-1)$ and weight $\bW(\tau-1)$. More specifically, the
evolution of $\sig(\tau)$ can be thought of as a time-varying Markov
chain with its transition matrix $P(\tau)$ being function of the time-varying
$\bW(\tau)$. That is, for $\Delta \geq 1$
\begin{align*}
\bmu(\tau) & =~\bmu(\tau - \Delta)P(\tau -\Delta)\cdots P(\tau-1).
\end{align*}
In above, we assume that the distribution $\bmu(\cdot)$ represents
an $|\cI(G)|$ dimensional row vector, $P(\cdot)$ represents  an
$|\cI(G)| \times |\cI(G)|$ probability transition matrix, and their
product on the right hand side should be treated as the usual
vector-matrix multiplication. The first step towards establishing
concentration of $\bmu(\tau)$ around the subset of $\cI(G)$ with
high-weight (cf.\ \eqref{eq:ov.4}) is establishing the existence of
an appropriate $\Delta \geq 1$:
\begin{itemize}
\item[(a)] $\Delta$ is small enough so that
\[
P( \tau-\Delta)\cdots P(\tau-1)\approx P(\tau)^{\Delta}.
\]

\item[(b)] $\Delta$ is large enough so that
\[
\bmu(\tau-\Delta)P(\tau)^{\Delta} \approx\bpi(\tau),
\]
where $\bpi(\tau)$ is the stationary distribution of $P(\tau)$,
i.e.\ $\bpi(\tau)=\bpi(\tau) P(\tau)$.
\end{itemize}
By finding such $\Delta$, it essentially follows that
$\bmu(\tau)\approx \bpi(\tau)$. The second step towards establishing
concentration of $\bmu(\tau)$ around the high-weight set involves
establishing that $\bpi(\tau)$ is approximately product-form with
respect to the weights $\bW(\tau)$ (cf.\ Lemma \ref{lem:pi}).
Therefore, as a consequence of Gibb\rq{}s maximal principle for
product-form distributions, it follows that $\bpi(\tau)$ is
concentrated around the subset of $\cI(G)$ with high-weight  (cf.\
\eqref{eq:ov.4}). Formally, this is stated in Proposition
\ref{prop:goodpi}. Subsequently, this establishes that $\bmu(\tau)$
is concentrated around the subset of $\cI(G)$ with high-weight (cf.\
\eqref{eq:ov.4})

Now we discuss the remaining task of showing the existence of
$\Delta$ so that (a) and (b) are satisfied. This is where we shall
discover another sets of conditions on $g$ that it must
be of the form $\exp(\log\log^\alpha x)$ with $\alpha>2$. 
Now for (b) to hold, it is required that $\Delta$ is larger than the
mixing time of $P(\tau)$. Using Cheeger\rq{}s inequality
\cite{DFK91,sinclair}, we prove that it is sufficient to have
\begin{align}\label{eq:ov.6}
 \Delta & > f_1(\bW_{\max}(\tau))\quad\mbox{with}\quad f_1(x)=\Theta(x^{6n+1}).
\end{align}
The precise definition of $f_1(\cdot)$ is presented in Lemma
\ref{lem:mixing}.\footnote{As noted in Section \ref{sec:not}, we use
the asymptotic notation $\Theta$ with respect to scaling in
$\bW_{\max}(\cdot)$
 instead of $n$.} Next, for
$\Delta$ to satisfy (a), observe that
\begin{align*}
 \|P( \tau-\Delta)\cdots P(\tau-1)-P(\tau)^{\Delta}\| & ~\leq~
\sum_{s=1}^{\Delta}\|P( \tau-\Delta)\cdots
P(\tau-s-1)(P(\tau-s)-P(\tau))P(\tau)^{s-1}\|\\ &~\leq~
\sum_{s=1}^{\Delta}\|P(\tau-s)-P(\tau)\|,
\end{align*}
where we use the triangle inequality with an appropriately defined
norm $\|\cdot\|$. Further, by exploring algebraic properties of
$P(\cdot)$ and $\bW(\cdot)$ (cf.\ Proposition \ref{prop:changep} and
\ref{clm:wbound}), we show that
$$\|P(\tau-s)-P(\tau)\|~\leq~
f_2(\bW_{\min}(\tau))\cdot s,$$ where $\bW_{\min}(\tau)=\min_i
W_i(\tau)$ and $f_2(x)=\Theta\left(\frac{x}{g^{(-1)}\left(\expo{\log^2
x}\right)}\right)$. Thus, it follows that
\begin{eqnarray*}
\|P( \tau-\Delta)\cdots P(\tau-1)-P(\tau)^{\Delta}\| &\leq&
f_2(\bW_{\min}(\tau))\cdot \Delta^2.
\end{eqnarray*}
Therefore, (a) follows if $\Delta$ satisfies
\begin{align}\label{eq:ov.7}
\Delta & <\frac{\varepsilon}{ \sqrt{f_2(\bW_{\min}(\tau))}} \quad \mbox{for small enough}~
\varepsilon>0.
\end{align}
From \eqref{eq:ov.6} and \eqref{eq:ov.7}, it follows that a $\Delta
\geq 1$ satisfying (a) and (b) exists if
\begin{align}\label{eq:ov.8}
f_1(\bW_{\max}(\tau)) & < \frac{\varepsilon}{\sqrt{f_2(\bW_{\min}(\tau))}}\quad\mbox{for large enough}~
\bQ_{\max}(\tau).
\end{align}
From \eqref{eq:weight}, it follows that for any $i \in V$,
\begin{align}
W_i(\tau) & ~\geq~\max_{j\in\mathcal N(i)} \exp\left({\sqrt{\log
g(A^i_j(\tau))}}\right)
              ~\approx~  \max_{j\in\mathcal N(i)} \exp\left({\sqrt{\log W_j(\tau)}}\right)
              ~ \geq~  \exp\left({\sqrt{\log W_j(\tau)}}\right),  \label{eq:ov.9}
\end{align}
for any $j \in \cN(i)$; here we have assumed $g(A_j^i(\tau)) \approx
W_j(\tau)$. Now let $j_* \in \arg\min_j W_j(\tau)$ and $j^* \in
\arg\max_j W_j(\tau)$. Since $G$ is connected, there exists a path
connecting $j_*$ and $j^*$ of length at most $D$ where $D\leq n-1$
is the diameter of $G$. Then by a repeated application of
\eqref{eq:ov.9} along this path joining $j_*$ and $j^*$ starting
with $j_*$, we obtain that
\begin{align}
\bW_{\min}(\tau) & \geq~ \exp\left(\log^{1/2^D} \bW_{\max}(\tau)\right). \label{eq:ov.10}
\end{align}
Therefore, the desired inequality \eqref{eq:ov.8} is satisfied for large $\bW_{\max}(\tau)$
if
\begin{equation*}
f_1(\bW_{\max}(\tau))< \frac{\varepsilon}{
\sqrt{f_2\left(\exp\left(\log^{1/2^D}
\bW_{\max}(\tau)\right)\right)}}.\end{equation*}
This holds if
\begin{eqnarray*}
\limsup_{x\to\infty} {f_1(x)}\sqrt{f_2\left(\exp\left(\log^{1/2^D}
x\right)\right)}~=~0. \end{eqnarray*} The above can be checked to
hold if $g$ does not grow
faster than $\exp(\log\log^{\alpha} x)$ for some constant  $\alpha<\infty$. 

\section{Preliminaries: Primary MC and Positive Recurrence}\label{sec:prelim}

\subsection{A Markov Chain (MC) of Interest}\label{sec:mcofint}
We describe a Markov chain of finite state space, whose time-varying
version will describe the evolution of the medium access algorithm
described in Section \ref{sec:algo}. As we described in Section
\ref{sec:pfsum}, our strategy for proving Theorem \ref{thm:main}
crucially relies on understanding the stationary distribution and
mixing time of the (finite state) Markov chain.
\paraG{Description}
The Markov chain evolves on state space $\cI(G)\times \{0,1\}^n$ and
uses node weights $\bW=[W_i] \in \Rp^n$ with $\bW_{\min} \geq 1$.
Given $(\sig, \ba) \in \cI(G) \times \{0,1\}^n$, the next (random)
state $(\sig^{\prime},\ba^{\prime})\in \cI(G) \times \{0,1\}^n$ is
obtained as follows:

\begin{itemize}
\item[1.] Each node $i$ chooses $r_i\in\{0,1\}$ uniformly at random,
i.e.\
$r_i = 1$ with probability $1/2$ and $0$ otherwise. Temporarily set
$$a^{\prime}_i ~=~
\begin{cases}
r_i &\mbox{if}~ a_j = 0~ \mbox{for all}~j \in \cN(i)\\
0&\mbox{otherwise}
\end{cases}.$$

\item[2.] Each node $i$ sets $\sigma^{\prime}_i$ (and possibly resets $a^{\prime}_i$)
as follows:
\begin{itemize}
\item[$\circ$] If $\sigma_i = 1$, then set
$$(\sigma^{\prime}_i,a^{\prime}_i) ~=~
\begin{cases}
(0,0)&\mbox{with probability}~\frac1 {W_i},\\
(1,1)&\mbox{otherwise.}
\end{cases}$$
\item[$\circ$] {Else if} $a_j = 0$ for all $j \in \cN(i)$, then set
$$\sigma^{\prime}_i =
\begin{cases}
1 & \mbox{if}~ a^{\prime}_i=1 ~\mbox{and}~ a^{\prime}_j = 0~\mbox{for all} ~j \in \cN(i) \\
0 & \mbox{otherwise}\end{cases}.$$
\item[$\circ$] Otherwise, set $(\sigma^{\prime}_i, a^{\prime}_i) = (0,0).$
\end{itemize}
\end{itemize}

\paraG{Stationary distribution}
Let $\Omega = \cI(G)\times \{0,1\}^n$. Then $\Omega$ is the state
space of the above described Markov chain; let $P_{\bx \bx^\prime}$
denote its transition probability for $\bx=(\sig,\ba),~
\bx^{\prime}=(\sig^{\prime}, \ba^{\prime}) \in \Omega$. We
characterize the stationary distribution of this Markov chain as
follows.
\begin{lemma}\label{lem:pi}
Staring from initial state $(\bzero,\bzero)$, the Markov chain $P$ is recurrent and aperiodic; let its
recurrence class be denoted by $\Omega^\prime \subset \Omega$; $(\sig, \bzero) \in \Omega^\prime$
for all $\sig \in \cI(G)$. Therefore, the Markov chain $P$ has a unique stationary distribution
$\bpi$ on $\Omega^\prime$  such that for any $(\sig,\ba)\in \Omega^\prime$
\begin{align}
\pi_{(\sig,\ba)} & \propto \exp\big({\sig\cdot\log \bW+U(\sig,\ba)}\big), \label{eq:statofP}
\end{align}
where $U: \Omega^\prime \to \Rp$ is such that $| U(\sig,\ba) |  \leq n 4^n \log 2$ for all $(\sig,\ba) \in \Omega^\prime$.
\end{lemma}
To achieve the form \eqref{eq:statofP}, we use the classical Markov
chain tree theorem \cite{AT89}. Our proof strategy can be of broad
interest to characterize such form for non-reversible Markov chains.
The proof of Lemma \ref{lem:pi} is presented in Appendix
\ref{sec:pflempi}.


\paraG{Mixing time} Now we establish a bound on the `mixing time' of
$P$ -- the time to reach near stationary distribution starting from
any initial distribution. We shall use the total-variation distance:
given distributions $\bnu, \bmu$ on a finite state space
$\Omega^\prime$, define
$\|\bnu - \bmu\|_{TV} = \sum_{\bx \in \Omega^\prime} |\nu_\bx -
\mu_\bx|.$
\begin{lemma}\label{lem:mixing}
Given $\varepsilon \in (0,0.5)$ with $n \geq 2$, for any distribution $\bmu$ on $\Omega^\prime$,
\begin{align*}
\norm{\bmu P^{\tau}-\bpi}_{TV} & < ~\varepsilon,
\end{align*}
for all $\tau\geq T_{{\mix}}(\varepsilon, n, \bW_{\max})$, where
\begin{align}\label{eq:tmix}
T_{\mix} \equiv T_{{\mix}}(\varepsilon, n, \bW_{\max}) & = 4^{n 4^{n+1} + 1} \bW_{\max}^{6n}~\log\Big(\frac{4^{n4^n} \bW_{\max}^{n}}{2\varepsilon}\Big).
\end{align}
\end{lemma}
We use the Cheeger's inequality \cite{DFK91, sinclair} to achieve
the mixing bound \eqref{eq:tmix}. The proof of Lemma
\ref{lem:mixing} is presented in Appendix \ref{sec:pflemmixing}.

\subsection{Ergodicity, Positive recurrence and Lyapunov-Foster}

To establish optimality of the medium access algorithm, we need to
show that the underlying network Markov chain, which has countably
infinite state space, is ergodic, i.e.\ that it has the unique
stationary distribution to which it converges. We briefly recall
known methods from literature that will be helpful in doing so.

Consider a discrete time Markov chain $X(\cdot)$ on countably infinite state
space $\sX$. State $\sx\in \sX$ is said to be {\em recurrent} if
$\Pr(T_\sx=\infty)=0,$  where $T_\sx = \inf\{\tau\geq 1 : X(\tau)=\sx : X(0)=\sx\}.$
Specifically, a recurrent state $\sx$ is called {\em positive recurrent} $\E[T_{\sx}]<\infty$,
or else if $\E[T_{\sx}] = \infty$ then it is called {\em null recurrent}. For an irreducible
Markov chain, if one of its state is positive recurrent, the so are all; we call such a Markov
chain positive recurrent.  An irreducible, aperiodic and positive recurrent Markov
chain is known to be ergodic: it has unique stationary and starting from any initial
distribution, it converges (in distribution) to stationary distribution. Therefore,
it is sufficient to establish positive recurrence property for establishing ergodicity
of the Markov chain in addition to verifying irreducibility and aperiodicity properties.
We shall recall a sufficient condition for establishing positive recurrence, known
as the Lyapunov and Foster's criteria.

\paraG{Lyapunov and Foster's criteria}
This criteria utilizes existence of a ``Lyapunov'', ``Potential'' or
``Energy'' function of the state under evolution of the Markov
chain. Specifically, consider a non-negative valued function  $L :
\sX \to \Rp$ such that $\sup_{\sx\in \sX} L(\sx) = \infty$. Let $h:
\sX \to \Zp$ be another function that is to be interpreted as a
state dependent ``stopping time''. The `drift' in Lyapunov function
$L$ in $h$-steps starting from $\sx \in \sX$ is defined as
\begin{align*}
\E[L(X(h(\sx)))-L(X(0))~|~X(0)=\sx\,]. 
\end{align*}
Following is the criteria (see \cite{Foss-Fluid}, for example):
\begin{theorem}\label{thm:PR}
For any $\kappa > 0$, let $B_{\kappa} = \{ \sx : L(\sx) \leq \kappa\}$.
Suppose there exist functions $h, k : \sX \to \Zp$ such that for any $\sx \in \sX$,
$$\E\left[L(X(h(\sx))) - L(X(0)) ~|~ X(0) = \sx \,\right] \leq -k(\sx),$$
that satisfy the following conditions:
\begin{itemize}
\item[(L1)] $\inf_{\sx \in \sX} k(\sx) > -\infty$.
\item[(L2)] $\lim\inf_{L(\sx) \to \infty} k(\sx) > 0$.
\item[(L3)] $\sup_{L(\sx) \leq \gamma} h(\sx) < \infty$
for all $\gamma > 0$.
\item[(L4)] $\lim\sup_{L(\sx)\to\infty} h(\sx)/k(\sx) < \infty$.
\end{itemize}
Then, there exists constant $\kappa_0 > 0$ so that for all
$\kappa_0 < \kappa$, the following holds:
\begin{eqnarray*}
\E\left[T_{B_\kappa}~|~X(0)=\sx \,\right] & < &  \infty, \qquad \mbox{for any $\sx \in \sX$} \label{eq:d1a} \\
\sup_{\sx \in B_\kappa} \E\left[ T_{B_\kappa}~|~X(0)=\sx\, \right] &
< & \infty, \label{eq:d2a}
\end{eqnarray*}
where $T_{B_\kappa}:=\inf\{\tau \geq 1:X(\tau)\in B_\kappa\}$ i.e.\
the first return time to $B_\kappa$. In other words, $B_\kappa$ is
positive recurrent.
\end{theorem}
Theorem \ref{thm:PR} implies that if {\it (L1) - (L4)} are satisfied and $B_\kappa$ is a finite set,
the Markov chain is positive recurrent.

\subsection{Notations}\label{sec:not}

Let $\Z$ and $\R$ ($\Z_+$ and $\R_+$) denote sets of (non-negative)
integers and reals, respectively.  Bold letters are reserved for
vector and distribution, e.g. $\bu = [u_i]$ denotes a vector;
$\bzero$ and $\bone$ represent vectors of all $0$'s and $1$'s; for a
function $f : \R \to \R$, we use $f(\bu)$ to denote $[f(u_i)]$.
Similarly for a random vector $\bu$, we use $\E[\bu]$ to denote
$[\E[u_i]]$. Let $\bu_{\max} := \max_i u_i$, $\bu_{\min} := \min_i
u_i.$ for a vector $\bu$; and $\bu \cdot \bv$ denote the inner
product $\sum_i u_i v_i$ of vectors $\bu, \bv$. We call $f : \R \to
\R$ as (uniformly) $c$-Lipschitz if $|f(x)-f(y)|\leq c |x-y|$ for
some constant $c > 0$. Similarly, a sequence of random variables
$\{A(\tau)\in\R:\tau \in \Z_+ \}$ is $c$-Lipschitz if
$|A(\tau)-A(\tau+1)|\leq c$ with probability $1$, for all $\tau\in
\Z_+$, for some constant $c > 0$.

We will use asymptotic notations (e.g.\ $O, o, \Omega, \omega,
\Theta$) with respect to scaling in queue-sizes, instead of the
network size or something else. For example, we mean $n=O(1)$ and
$2^n\bQ_{\max}=O(\bQ_{\max})$ where $n$ is the number of nodes (or
queues). We say function $f :\R_+\to \R_+$ is polynomial by denoting
$f(x)={\bf poly}(x)$ if $\lim_{x\to\infty}\frac{f(x)}{x^c}=0$ for
some constant $c>0$. Similarly, $f(x)={\bf superpoly}(x)$ and
$f(x)={\bf superpolylog}(x)$ mean
$\lim_{x\to\infty}\frac{f(x)}{x^c}=\infty$ and
$\lim_{x\to\infty}\frac{f(x)}{\log^c x}=\infty$ for any constant
$c>0$, respectively.

\section{Proof of Theorem \ref{thm:main}}\label{sec:proof}

We shall establish ergodicity of an appropriate Markov chain
describing evolution of the network under medium access algorithm as
long as $\lamb \in \Lamb$. To that end, recall the Markov state of
the network defined in the Section \ref{sec:pfsum}:
\[
X(\tau) = (\bQ(\tau), \sig(\tau),\ba(\tau), \bA(\tau), \bB(\tau)),
\]
where recall that $\bQ(\tau)$ represents vector of queue-sizes; $\ba(\tau) \in \{0,1\}^n$
denotes the vector of transmission attempts by nodes at time $\tau$;
$\sig(\tau) \in \cI(G)$ denotes the vector of resulting successful
transmissions in time $\tau$ (clearly, $\sig(\tau) \leq \ba(\tau)$); and
$\bA(\tau), ~\bB(\tau) \in \Zp^{2|E|}$ denoting the vector of
long-term and short-term estimations maintained at nodes of the
weights of their neighbors as explained in Section \ref{sec:algo}.
Then it follows that under medium access algorithm
$X(\cdot)$ is a Markov chain. It can be easily checked that under this
Markov chain, state $\bzero$ in which all components are $0$, has
positive probability of transiting to itself. Further, starting from
any state, $X(\cdot)$  has positive probability of reaching state
$\bzero$. Therefore, $X(\cdot)$ is always restricted to the recurrence
class containing state $\bzero$; and over this class it is aperiodic.
Therefore, it is sufficient to establish positive recurrence of $X(\cdot)$
over this recurrence class. Towards this, we shall utilize the
following Lyapunov function $L$ and auxiliary functions $h, k$
to verify the conditions of Theorem \ref{thm:PR}. Given state
$\sx = (\bQ,\sig,\ba,\bA, \bB)$, define
\begin{align}\label{eq:def.lyp}
L(\sx) & = \sum_i F(Q_i) + \sum_{i; j \in \cN(i)} \Big((A_j^i)^2 + g^{(-1)}(B^i_j) \Big),
\end{align}
where $F(x) = \int_0^x \log\log y~dy$ with $\log \log y = [\log\log
y]_+$; let $g^{(-1)}(x) = \exp(\exp(\log^{1/4} x))$ represent the
inverse function of $g(x) = \exp(\log\log^4 x)$. With an abuse of
notation, we shall use $L(\tau)$ to denote $L(X(\tau))$.

Recall that node weights $\bW$ are determined by $\bQ$ and $\bA$ as per \eqref{eq:weight}.
Therefore, given state $\sx = (\bQ,\sig,\ba,\bA, \bB)$, the weight vector $\bW$ is
determined. With this in mind, let
\begin{align}
C(\sx) & = \max \big\{ g(\bA_{\max}), ~\bB_{\max}\big\}.
\end{align}
Then $h$ and $k$ are defined as
\begin{align}
h(\sx) & =\left\{
\begin{array}
[c]{cc}%
C(\sx)^n & \mbox{if}~C(\sx) \geq \bW_{\max}^3, \\
                                         \frac{1}{2} \exp\big(\exp(\log^{1/2} \bW_{\max})\big) & \text{otherwise.$\qquad\quad$}
                 \end{array}\right.\label{eq:def.h}\\
k(\sx) & = \left\{
\begin{array}
[c]{cc}
C(\sx)^{2n} & \mbox{if}~C(\sx) \geq \bW_{\max}^3, \\
                                         \frac{\log^{1/2} \bW_{\max}}{2} \exp\big(\exp(\log^{1/2} \bW_{\max})\big) &
                                         \text{otherwise.$\qquad\quad$}
                 \end{array}\right.\label{eq:def.k}
\end{align}
With these definitions, we shall establish the following.
\begin{lemma}\label{lem:PR}
Let $\lamb \in \Lamb$. Then for any $\sx$ with $L(\sx)$ large enough,
\begin{align}
\E\big[L(h(\sx)) - L(0) \,|\, X(0) = \sx\big] & \leq - k(\sx).
\end{align}
\end{lemma}
It can be easily checked that $L, h$ and $k$ along with Lemma
\ref{lem:PR} satisfy conditions of Theorem \ref{thm:PR}. Now $L(\sx)
\to \infty$ as $|\sx| \to \infty$ where $|\sx|= |\bQ| + |\sig| +
|\ba| + |\bA| + |\bB|$ with $|\sig|, ~|\ba|$ being equal to the
ordering of them and $|\bQ|, |\bA|$ and $|\bB|$ are standard
$1$-norm. Therefore, $B_\kappa = \{ \sx : L(\sx) \leq \kappa\}$ is a
finite set. Therefore, it follows that the Markov chain $X(\cdot)$
is positive recurrent; it is aperiodic and irreducible on the
recurrence class containing $\bzero$ as discussed before. Therefore,
it follows that it is ergodic. That is medium access algorithm of
interest is optimal establishing Theorem \ref{thm:main}. In the
remainder this section, we shall establish the key Lemma
\ref{lem:PR}. As explained in Section \ref{sec:pfsum}, the
proof is divided in two cases: (a) for $\sx$ with $C(\sx) \geq \bW_{\max}^3$
and (b) otherwise.

The case (a) corresponds to the situation when at least one of the
estimation $g(A_j^i(\cdot)), B_j^i(\cdot)$ of $W_j(\cdot)$  some neighbor $j \in \cN(i)$
for some node $i$ is quite large. Therefore, in this case, due to the nature of the
algorithm, we show that there is a reduction in the Lyapunov function
(part that depends on $\bA(\cdot), ~\bB(\cdot)$). This is argued in detail
in Section \ref{sec:proof.a}.

In case (b), on the other hand, all estimations are not too large.
Therefore, effectively the algorithm acts as if weight of each node, say node
$i$, is such that
\[
W_i(\cdot) \approx \max\Big\{\log Q_i(\cdot), \max_{j\in \cN(i)} \exp\Big(\sqrt{\log W_j(\cdot)}\Big)\Big\}.
\]
Given this, as long as the $\bW_{\max}$ (equivalently $\bQ_{\max}$) is large enough,
weight of each node is large enough (as it can be shown to be lower bounded by some
increasing function of $\bQ_{\max}$). Therefore, weight of each node changes very slowly :
each component of $\bQ(\cdot)$ changes at most by unit per unit time and hence if
$\bQ_{\max}$ is large then $\log \bQ_{\max}$ changes by small amount per unit time.
This essentially `freezes' the weights over a time period that is long enough for the
corresponding Markov chain of $(\sig(\cdot), \ba(\cdot))$ to reach its stationary
distribution (using bound on Mixing time cf. Lemma \ref{lem:mixing}). We show that
the stationary distribution has property that (with respect to it) on average
the first part of the Lyapunov function decreases maximally; it results into overall
negative drift if $\lamb \in \Lamb$. This will be useful to conclude Lemma \ref{lem:PR} in
case (b). This is argued in detail in Section \ref{sec:proof.b}.

\section{Proof of Lemma \ref{lem:PR}: $C(\sx) \geq
\bW_{\max}^3$}\label{sec:proof.a}


The goal is to establish that starting with state $X(0) = \sx = (\bQ,\sig,\ba,\bA,\bB)$
such that $L(\sx)$ is large enough (to be determined in the course of the proof) and
$C(\sx) = \max\{g(\bA_{\max}), \bB_{\max}\}$ $\geq \bW_{\max}^3$ (with $\bW$
determined based on $\bQ,~\bA$ as per \eqref{eq:weight}), after time
$h(\sx) = C(\sx)^n$ the expected value of $L$ decreases by $k(\sx) = h(\sx)^2 = C(\sx)^{2n}$.
\begin{align}\label{eq:lem1}
\E\big[L(h(\sx)) - L(0) \,|\, X(0) = \sx\big] & \leq - k(\sx).
\end{align}
We note that for proving Lemma \ref{lem:PR}, these will be the
definition of functions $h$ and $k$ as it concerns the case $C(\sx)
\geq \bW_{\max}^3$. To simplify notations, we will use notation
$\E[\cdot]$ and $\Pr[\cdot]$ instead of $\E[\cdot\,|\,X(0) = \sx]$
and $\Pr[\cdot\,|\, X(0)=\sx]$ whenever it is clear from the context.

To that end, note that if $L(\sx)$ is large enough, then either $\bQ_{\max}$,
$\bA_{\max}$ or $\bB_{\max}$ is large. Now if $\bQ_{\max}$ or $\bA_{\max}$
are large then necessarily $\bW_{\max}$ is large. Since $C(\sx)$ depends
on $\bA_{\max}, ~\bB_{\max}$ and since we have $C(\sx) \geq \bW_{\max}^3$,
it necessarily follows that $C(\sx)$ is large due to $L(\sx)$ being large.
Now for large enough  $L(\sx)$ and hence large enough $C(\sx)$,
\begin{align*}
h(\sx) & = C(\sx)^n ~\leq~ \frac{g^{(-1)}\Big(\exp\big(\log^2 (\sqrt{C(\sx)/2}-1)\big)\Big)}{2\big(\sqrt{C(\sx)/2}-1\big)}.
\end{align*}
The above holds for large enough $C(\sx)$ because
$g^{(-1)}(\exp(\log^2 x))$ is a super-polynomial function of $x$,
i.e.\
$$ \frac{g^{(-1)}(\exp(\log^2 x))}{x^c} \to \infty, \quad \mbox{as}~x\to\infty, \quad\mbox{for any fixed~} c > 0.$$
Therefore, from
\eqref{eq:cor1.a} of Corollary \ref{cor1:wbound} (presented in Appendix), it follows that for $\tau\leq h(\sx)$,
\begin{align}
\bW_{\max}(\tau) & \leq \sqrt{C(\sx)/2} ~\stackrel{\triangle}{=} \cW_{\max}. \label{eq:wmax}
\end{align}


\paraG{Two Lemmas} Now we state two key lemmas that will lead to \eqref{eq:lem1}. We shall present
their proofs in Section \ref{sec:pflem2} and \ref{sec:pflem2-1} respectively.
\begin{lemma}\label{lem2}
Given initial state $X(0) = \sx =(\bQ,\sig,\ba,\bA,\bB)$, let
$C(\sx) \geq \bW_{\max}^3$ and $C(\sx)$ be large enough. Then for
any $i$ and $j \in \cN(i)$
\begin{align}\label{eq:lem.2}
\E\big[A^i_j(h(\sx))^2\big] &\leq
\begin{cases}
(A^i_j)^2- \frac{A^i_j\, h(\sx)}{O\Big(g(A^i_j)^{\frac{n+1}2}\Big)}
+O\big(A^i_j\big)& \mbox{if}~
g(A^i_j)>\frac{C(\sx)}{2},\\
(A^i_j+ h(\sx))^2 & \mbox{otherwise}.
\end{cases}
\end{align}
\end{lemma}
\begin{lemma}\label{lem2-1}
Given initial state $X(0) = \sx=(\bQ,\sig,\ba,\bA,\bB)$, let $C(\sx)
\geq \bW_{\max}^3$ and $C(\sx)$ be large enough. Then for any $i$
and $j \in \cN(i)$
\begin{align}\label{eq:lem.2-1}
\E\big[g^{(-1)}(B^i_j(h(\sx))) \big]&\leq   O\Big(g^{(-1)}\big(C(\sx)/2\big)\Big).
\end{align}
\end{lemma}

\paraG{Implications of Lemmas \ref{lem2} and \ref{lem2-1}}
Define five different events as follows:
\begin{align*}
S_1  & =  \{ (i,j) \in E : g(A^{i}_j)=C(\sx) \} \\
S_2& =  \{(i,j) \in E : C(\sx) /2 < g(A^{i}_j) < C(\sx)\} \\
S_3&= \{(i,j) \in E : g(A^{i}_j) \leq C(\sx)/2\}\\
S_4& = \{(i,j) \in E: B^i_j= C(\sx)\}\\
S_5& = \{(i,j) \in E : B^i_j< C(\sx)\}.
\end{align*}

For $(i,j)\in S_1$, using relation $g(A_j^i) = C(\sx)$, $h(\sx) =
C(\sx)^n$ and Lemma \ref{lem2}, we have
\begin{align}
\E\big[A^i_j(h(\sx))^2 - (A^i_j)^2\big] &\leq
- \frac{A^i_j\, h(\sx)}{O\Big(g(A^i_j)^{\frac{n+1}2}\Big)}+O(A^i_j) \notag\\
&= - \frac{g^{(-1)}(C(\sx)) C(\sx)^n}{O\Big(C(\sx)^{\frac{n+1}2}\Big)}  + O\Big(g^{(-1)}(C(\sx))\Big)\notag\\
&\leq- \frac{1}{2}g^{(-1)}(C(\sx)),\label{eq:S1}
\end{align}
where the last inequality follows for large enough $C(\sx)$.

For $(i,j)\in S_2$, it follows from Lemma \ref{lem2} that for large
enough value of $C(\sx)$
\begin{align}\label{eq:S2}
\E\big[A^i_j(h(\sx))^2 - (A^i_j)^2\big] &\leq 0,
\end{align}
where we use $g(A_j^i)^{\frac{n+1}{2}}  = o\left(C(\sx)^n\right)$
and $g(A_j^i)=\Omega(C(\sx))$.

For $(i,j)\in S_3$, Lemma \ref{lem2} implies that
\begin{align}
&\E\big[A^i_j(h(\sx))^2-(A^i_j)^2\big]  ~\leq
~\big(A^i_j+h(\sx))^2-(A^i_j)^2~ =~ 2A^i_j\,h(\sx) + h(\sx)^2\notag\\
&\qquad\leq~  2g^{(-1)}\Big(\frac{C(\sx)}2\Big) h(\sx) + h(\sx)^2
~=~
O\Big(g^{(-1)}\Big(\frac{C(\sx)}2\Big)\Big)C(\sx)^{n},\label{eq:S3}
\end{align}
where the last inequality utilizes the super-polynomial property of
$g^{(-1)}(\cdot)$ function:
\begin{align}
h(\sx) = C(\sx)^{n} & = o\Big( g^{(-1)}\Big(\frac{C(\sx)}2\Big)\Big).
\end{align}

For $(i,j)\in S_4$, Lemma \ref{lem2-1} implies that
\begin{align}
\E\big[g^{(-1)}(B^i_j(h(\sx)))-g^{(-1)}(B^i_j)\big] &\leq
O\Big(g^{(-1)}\Big(\frac{C(\sx)}2\Big)\Big) - g^{(-1)}\Big(C(\sx)\Big) \notag\\
&\leq -\frac{1}{2} g^{(-1)}\Big(C(\sx)\Big),\label{eq:S4}
\end{align}
where the last equality follows for $C(\sx)$ large enough from the
following proposition stating the super-polynomial property of
$g^{(-1)}(\cdot)$ function; we skip the proof as it is elementary.
\begin{proposition}\label{clm:condg}
For any given $k \in \mathbb{Z}_+$,
\begin{align}
\lim_{x\to\infty} \frac{x^k g^{(-1)}(x/2)}{g^{(-1)}(x)}  & = 0.
\end{align}
\end{proposition}

For $(i,j)\in S_5$, Lemma \ref{lem2-1} implies that
\begin{align}
\E\big[g^{(-1)}\big(B^i_j(h(\sx))\big)\big]& = O\Big(g^{(-1)}\Big(\frac{C(\sx)}2\Big)\Big).\label{eq:S5}
\end{align}

From \eqref{eq:S1}, \eqref{eq:S2}, \eqref{eq:S3}, \eqref{eq:S4} and \eqref{eq:S5},
it follows that
\begin{align}
&\E\Big[\sum_{i,j} A^i_j(h(\sx))^2 + \sum_{i,j}
g^{(-1)}\big(B^i_j(h(\sx))\big)\Big]
    - \sum_{i,j} (A^i_j)^2 - \sum_{i,j} g^{(-1)}\big(B^i_j\big)\notag\\
& \qquad \leq - \Big(\frac{|S_1|+|S_4|}{2}\Big) g^{(-1)}(C(\sx)) +
(|S_3|+|S_5|) O\Big(g^{(-1)}\Big(\frac{C(\sx)}2\Big)\Big) C(\sx)^{n}\notag \\
& \qquad \stackrel{(a)}{=} -\frac{1}{2}  g^{(-1)}(C(\sx))  +
O\Big(n^2 ~g^{(-1)}\Big(\frac{C(\sx)}2\Big)\Big) C(\sx)^{n}\notag\\
& \qquad \stackrel{(b)}{\leq} -\frac{1}{4} g^{(-1)}(C(\sx)),\notag
\end{align}where (a) is from $|S_1|+|S_4|\geq 1$ and $|S_3| +
|S_5| \leq |E| \leq n^2$; (b) is from Proposition \ref{clm:condg}.

\paraG{Concluding \eqref{eq:lem1}}
From the above inequality, we obtain the following:
\begin{align*}
&\E\Big[L(h(\sx))-L(0)\Big]  \\
&\qquad=
\E\Big[\sum_i F(Q_i(h(\sx))) + \sum_{i,j} A^i_j(h(\sx))^2 + \sum_{i,j} g^{(-1)}\big(B^i_j(h(\sx))\big)\Big]\\
&\qquad \qquad-\E\Big[\sum_i F(Q_i) + \sum_{i,j} (A^i_j)^2+
\sum_{i,j} g^{(-1)}\big(B^i_j\big)\Big]\\
&\qquad{\leq} \E\Big[\sum_i F(Q_i(h(\sx))) - F(Q_i)\Big] - \frac{1}{4} g^{(-1)}(C(\sx))\\
&\qquad\stackrel{(c)}{\leq} \sum_i \big(F(Q_i+h(\sx))- F(Q_i)\big) -
\frac{1}{4} g^{(-1)}(C(\sx))\\
&\qquad\leq \sum_i f(Q_i+h(\sx))h(\sx) -\frac{1}{4} g^{(-1)}(C(\sx))\\
& \qquad\leq n f(\bQ_{\max}+h(\sx))h(\sx) - \frac{1}{4} g^{(-1)}(C(\sx)) \\
&\qquad\leq  n f\big(\exp(\bW_{\max}) + h(\sx)\big) h(\sx) - \frac{1}{4} g^{(-1)}(C(\sx)) \\
&\qquad\stackrel{(d)}{\leq}  n f\big(\exp(C(\sx)^{1/3}) + C(\sx)^n \big) C(\sx)^n - \frac{1}{4} g^{(-1)}(C(\sx))\\
&\qquad\stackrel{(e)}{=} - \frac{1}{8} g^{(-1)}(C(\sx))
~\stackrel{(f)}{\leq} - C(\sx)^{2n},
\end{align*}
where (c) is from 1-Lipschitz property of $Q_i(\cdot)$; (d) is from
$C(\sx) \geq \bW_{\max}^3$; (e) and (f) hold for large enough
$C(\sx)$ due to the fact that $f(x) = \log\log x$ and $g^{(-1)}(x)$
has the super-polynomial growth property as per Proposition \ref{clm:condg}.
This completes the proof of Lemma  \ref{lem:PR} for the case $C(\sx) \geq \bW_{\max}^3$.

\subsection{Proof of Lemma \ref{lem2}}\label{sec:pflem2}

Observe that Lemma \ref{lem2} for the case $g(A^i_j)\leq C(\sx)/2$
follows immediately from the 1-Lipschitz property of $A^i_j(\cdot)$.
Hence, we shall only consider the case when $g(A^i_j)> C(\sx)/2$.

Define $\cW_{\max} = \sqrt{C(\sx)/2}$. Then from \eqref{eq:wmax}
\begin{align}
\bW_{\max}(\tau) & < \cW_{\max} <  \sqrt{g(A^i_j)},\qquad\mbox{for
all}~~\tau\leq h(\sx). \label{eq5}
\end{align}
For the purpose of the proof, consider modification of the original
network Markov chain, say original Markov chain be $M$ and
let its modification be $M^{\prime}$. Under this modification, the $M^{\prime}$
evolves in the same manner as $M$ for $\tau \leq h(\sx)$;
for $\tau > h(\sx)$ the evolution of $M^{\prime}$ is the same as
that of $M$ except $\bW(\tau)=\bW(\tau-1)$ i.e.\ $\bW(\cdot)$ remains
fixed to its value at time $h(\sx)$. Clearly, the quantity of interest
$\E[A^i_j(h(\sx))^2]$ in Lemma \ref{lem2} remains invariant with
respect to $M$ and $M^{\prime}$. As mentioned earlier, this
modification is for convenience of proof and it merely guarantees
\eqref{eq5} for all $\tau$. Therefore, we shall bound $\E[A^i_j(h(\sx))^2]$
under $M^{\prime}$ for which we have
\begin{align}
\bW_{\max}(\tau) & < \cW_{\max} < ~\sqrt{g(A^i_j)},\qquad\mbox{for
all}~~\tau\geq 0\label{eq5-1}.
\end{align}
With respect to $M^{\prime}$, define random times $0 = T_0 < T_1 <
T_2 \dots$ such that $T_m$ is the $m$th time when $A^i_j(\cdot)$ is
updated, i.e.\ $B^i_j(T_m-1) \geq 2$ and $B^i_j(T_m)=0$. Define for
$m \geq 0$,
\begin{align}
Y_m & = \begin{cases}
A^i_j(T_m)^2 & \mbox{if}~ T_{m-1}\leq h(\sx)~~ \mbox{or}~~m=0 \\
Y_{m-1}- A^i_j& \mbox{otherwise}.
\end{cases}
\end{align}
Let $m^* = \inf\{m \geq 0: T_m > h(\sx)\}$. Then it follows that
\begin{align}
A^i_j(h(\sx))^2 & = A^i_j(T_{m^*-1})^2 ~=~Y_{m^*-1}.
\end{align}
We establish the following property of $Y_m$.
\begin{proposition}\label{clm3}
Given $g(A^i_j)> C(\sx)/2$, for $m \geq 1$
\begin{equation*}
\E[Y_{m+1}~|~\mathcal{F}_m]\leq Y_m- A^i_j,
\end{equation*}
where $\mathcal{F}_m$ denotes the filtration containing
$Y_k, T_k$ for $0\leq k\leq m$.
\end{proposition}
\begin{proof}
If $T_m > h(\sx)$, then the desired result follows from definition
of $Y_m$. Now suppose $T_m \leq h(\sx)$. Then observe that
\begin{align}
g(A^i_j(T_m)) &\stackrel{(a)}{\geq}  ~g(A^i_j-h(\sx))\notag\\
&\stackrel{(b)}{\geq} g(A^i_j)-h(\sx) g^{\prime}(c), \qquad\mbox{for some}~c\in (A^i_j-h(\sx),A^i_j)\notag\\
&\geq g(A^i_j)-h(\sx) g^{\prime}(A^i_j-h(\sx))\notag\\
&\stackrel{(c)}{\geq} g(A^i_j) - 1,\label{eq1-clm3}
\end{align}
where (a) is from 1-Lipschitz property of $A^i_j(\cdot)$; (b) is
from the mean value theorem; for (c) we use the following that holds
for large enough $C(\sx)$ and $g(A^i_j) \geq C(\sx)/2$ (along with
the definition of $g(\cdot)$)
\[ h(\sx) = C(\sx)^n \leq 2^{-n} g(A^i_j)^n ~ < ~ \sqrt{A^i_j} \]
and $g^{\prime}(x)<1/\sqrt{x}$ for large enough $x$.

Now we bound the probability that $A^i_j$ increases at time
$T_{m+1}$:
\begin{align}
\Pr\Big(A^i_j(T_{m+1}) = A^i_j(T_m)+1~\Big|~\mathcal{F}_m\Big) & =
\Pr\Big(B^i_j(T_{m+1}-1)\geq g(A^i_j(T_m))~\Big|~\mathcal{F}_m\Big) \notag\\
&\stackrel{(a)}{\leq}  \Pr\Big(B^i_j(T_{m+1}-1) \geq g(A^i_j)-1~\Big|~\mathcal{F}_m\Big)\notag\\
&\stackrel{(b)}{<} \Big(1-\frac{1}{\cW_{\max}}\Big)^{g(A^i_j)-2}\notag\\
&\stackrel{(c)}{<} \Big(1-\frac{1}{\sqrt{g(A^i_j)}}\Big)^{g(A^i_j)-2}\notag\\
&\stackrel{(d)}{\leq}  \frac{1}{10}.\label{eq2-clm3}
\end{align}
In above, (a) and (c) are from \eqref{eq1-clm3} and \eqref{eq5-1}
when assuming $A^i_j$ (equivalently $C(\sx)$) is large enough. For
(b), we observe that $\bW_{\max}(\tau)$ is uniformly bounded above
by $\cW_{\max}$ from (\ref{eq5-1}). Therefore, once $j$ is
successful in its transmission, the probability that $j$
consecutively attempts to transmit (without stopping) for an
interval of length $k$ is at most
$\left(1-\frac1{\cW_{\max}}\right)^{k}$. Using \eqref{eq2-clm3}, it
follows that
\begin{align*}
\E\Big[Y_{m+1} \Big| \mathcal{F}_m\Big] & =  \E\Big[A^i_j(T_{m+1})^2\Big|\mathcal{F}_m\Big] \\
&\leq  \frac1{10}\Big(A^i_j(T_{m})+1\Big)^2 + \frac9{10}\Big(A^i_j(T_{m})-1\Big)^2\\
&= A^i_j(T_{m})^2 - \frac85 A^i_j(T_{m})\\
&\leq A^i_j(T_{m})^2- \frac85 A^i_j + \frac85 T_m\\
&\leq A^i_j(T_{m})^2-\frac85 A^i_j  + \frac85 h(\sx) \\
&\leq Y_m-A^i_j,
\end{align*}
where we used 1-Lipschitz property of $A^i_j(\cdot)$, $T_m \leq
h(\sx)$ and the fact that $h(\sx) = C(\sx)^n \leq 2^{-n} g(A_j^i)^n
= o(A^i_j)$. This completes the proof of Proposition \ref{clm3}.
\end{proof}

\paraG{Completing proof of Lemma \ref{lem2}}
Define $Z_m = Y_m + (m-1)A^i_j$. Then as per Proposition \ref{clm3},
$\{ Z_m: m\geq 1\}$ is a sub-martingale with respect to
$\mathcal{F}_m$. By the Doob's optional stopping theorem, we have
that
\[
\E\big[Z_{m^*}\big] \leq \E[Z_1] = \E[Y_1].
\]
Therefore, the desired inequality follows as
\begin{align*}
\E\Big[A^i_j(h(\sx))^2]& = \E\Big[A^i_j(T_{m^*-1})^2\Big]
~\stackrel{(a)}{\leq}~ \E\Big[(A^i_j(T_{m^*})+1)^2\Big]\\
&= \E\Big[A^i_j(T_{m^*})^2\Big] + 2 \E\Big[A^i_j(T_{m^*})\Big] + 1
\end{align*}
\begin{align*}
&\stackrel{(b)}{\leq} \E\Big[Y_{m^*}\Big] + 2 \E\Big[A^i_j+m^*\Big] + 1\\
&= \E\Big[Z_{m^*}-(m^*-1)A^i_j\Big] + 2 \E\Big[A^i_j + m^*\Big] + 1\\
&\leq \E\Big[Y_1\Big] - \E\Big[m^*\Big] \big(A^i_j-2\big) + 3 A^i_j + 1\\
&=\E\Big[A^i_j(T_1)^2\Big] - \E\Big[m^*\Big]\big(A^i_j-2\big) + 3 A^i_j + 1\\
&\stackrel{(c)}{\leq} \big(A^i_j+1\big)^2 - \E\Big[m^*\Big] \big(A^i_j-2\big) + 3 A^i_j + 1\\
&= (A^i_j)^2  -  \E\Big[m^*\Big]\big(A^i_j-2\big) + 5 A^i_j + 2\\
&\stackrel{(d)}{=} (A^i_j)^2-
\frac{h(\sx)}{O\Big(g(A^i_j)^{\frac{n+1}2}\Big)}  A^i_j +
O\big(A^i_j\big),
\end{align*}
where (a), (b), (c) are from the 1-Lipschitz property of $A^i_j(\cdot)$ and
(d) is due to the following proposition. This completes the proof of Lemma \ref{lem2}.

\begin{proposition}\label{clm5}
For large enough $C(\sx)$, $$\E[m^*]~\geq~
\frac{h(\sx)}{O\left(g(A^i_j)^{\frac{n+1}2}\right)}+1.$$
\end{proposition}
\begin{proof}
For $1\leq \tau\leq h(\sx)$, define
\[
U_{\tau} =  \begin{cases}
1 &\mbox{if $A^i_j(\cdot)$ is updated at time $\tau$}\\
0 &\mbox{otherwise}.
\end{cases}
\]
That is, $U_{\tau}=1$ iff $B^i_j(\tau-1)\geq 2$ and $B^i_j(\tau)=0$. By definition
of $U_{\tau}$ and $m^*$, $$m^*-1=\sum_{\tau=1}^{h(\sx)} U_{\tau}.$$ Therefore,
to bound $\E[m^*]$ we next bound $\E[U_\tau]$. For any $5\leq \tau \leq h(\sx)-5$,
let $X(\tau-5)=\{\bQ(\tau-5), \sig(\tau-5), \ba(\tau-5), \bA(\tau-5), \bB(\tau-5)\}$
be the network state at time $\tau-5$. For this, define event
$\mathfrak{E}$:
\begin{eqnarray*}
\mathfrak{E}&=&\mathfrak{E}^{\prime}_1~\&~\mathfrak{E}^{\prime}_2~\&~\mathfrak{E}^{\prime}_3\\
\mathfrak{E}_1&=&\mbox{all nodes do not attempt to transmit at time $\tau-4$}\\
\mathfrak{E}_2&=&\mbox{Only $j$ attemtps to trasmit at time $\tau-2$ and $\tau-3$}\\
\mathfrak{E}_3&=&\mbox{$j$ does not attempt to transmit at time
$\tau-1$}.
\end{eqnarray*}
If $\mathfrak{E}$ happens, $A^i_j$ is updated at time $\tau$ i.e.\
$U_{\tau}=1$. First note that
\begin{equation}\label{eq1-clm5}
\Pr[\mathfrak{E}_1]\geq
\Big(\frac1{\cW_{\max}}\Big)^n=\Omega\Big(\frac1{g(A^i_j)^{n/2}}\Big),
\end{equation}
whether this naive lower bound is obtained from \eqref{eq5-1}
and the case when many nodes (as possible) succeed in their transmissions at time $\tau-4$.
Second we have
\begin{equation}\label{eq2-clm5}
\Pr[\mathfrak{E}_2~|~\mathfrak{E}_1]\geq \left(\frac12\right)^n\times \left(\frac12\right)^n=\Omega(1).
\end{equation}
The above lower bound is obtained considering the scenario that the balanced coin
of $j$ produces `head' at times $\tau-2$ and $\tau-3$; coins of all other nodes
produce 'tail' at times $\tau-2$ and $\tau-3$. Third since the transmission
of $j$ is successful at time $\tau-2$, it is easy to see that
\begin{equation}\label{eq3-clm5}
\Pr[\mathfrak{E}_3~|~\mathfrak{E}_2]\geq
\frac1{\cW_{\max}}=\Omega\Big(1/{\sqrt{g(A^i_j)}}\Big),
\end{equation}
from \eqref{eq5-1}.
By combining \eqref{eq1-clm5}, \eqref{eq2-clm5} and \eqref{eq3-clm5},
\begin{align*}
&\Pr[U_{\tau}=1~|~X(\tau-5)]\\
&\qquad\geq~ \Pr[\mathfrak{E}~|~X(\tau-5)]\\
&\qquad=~\Pr[\mathfrak{E}_1~\&~\mathfrak{E}_2~\&~\mathfrak{E}_3~|~X(\tau-5)]\\
&\qquad=~ \Omega\left(g(A^i_j)^{-\frac{n+1}2}\right).
\end{align*}
The above inequality holds for any given $X(\tau-5)$. Hence,
$$\Pr[U_{\tau}=1]~=~ \Omega\left(g(A^i_j)^{-\frac{n+1}2}\right).$$
Finally, the conclusion follows as
\begin{align*}
\E[m^*-1]&~\geq~ \E\left[\sum_{\tau=5}^{h(\sx)} U_{\tau}\right]\\& ~=~
\sum_{\tau=5}^{h(\sx)} \E[U_{\tau}]\\
&~=~ \sum_{\tau=5}^{h(\sx)} \Pr[U_{\tau}=1]\\&~=~ (h(\sx)-4)\cdot
\Omega\left(g(A^i_j)^{-\frac{n+1}2}\right)\\& ~=~
\frac{h(\sx)}{O\left(g(A^i_j)^{\frac{n+1}2}\right)}.
\end{align*}
\end{proof}

\subsection{Proof of Lemma \ref{lem2-1}}\label{sec:pflem2-1}

Let $\tau^*=\inf\{\tau \geq 1:a_j(\tau)=0\}$ i.e.\ the
first time $j$ does not attempt to transmit, and let the event
$\mathfrak{E}$ denote $\tau^*\geq h(\sx)$. Hence, if $\mathfrak{E}$
happens, $B^i_j(h(\sx))=B^i_j+h(\sx)$ and transmissions of $j$ should be
successful consecutively for time $\tau\in [0,h(\sx)-2]$ (otherwise,
$j$ would have stopped attempting). Under this observation, we obtain
\begin{eqnarray}
\Pr[\mathfrak{E}]&\leq&\Pr\left[\mbox{$j$ attempts to transmit consecutively for time $\tau\in [1,h(\sx)-1]$}~\right]\notag\\
&\leq&\Big(1-\frac1{\cW_{\max}}\Big)^{h(\sx)-1},\label{eq6}
\end{eqnarray}
where the last inequality follows from the fact that $W_j(\tau)$ is bounded from above by
$\cW_{\max}$ as per \eqref{eq:wmax}. On the other hand, if the
event $\mathfrak{E}$ does not happen, $j$ stops attempting transmission
before time $h(\sx)$. Hence $B^i_j$ should set to $0$ before time $h(\sx)$.
Based on this observation and arguments similar to those used for establishing \eqref{eq6},
we obtain
\begin{eqnarray}
\Pr[B^i_j(h(\sx))=k~|~\mathfrak{E}^c]
&\leq&\Pr\Big[\mbox{$j$ attempts to transmit consecutively}\notag\\
&&\qquad\qquad \mbox{for time $\tau\in [h(\sx)-k+1,h(\sx)-1]$}~\Big]\notag\\
&\leq&\Big(1-\frac{1}{\cW_{\max}}\Big)^{k-1}.\label{eq7}
\end{eqnarray}
Now observe that
\begin{align}
\E\left[g^{(-1)}(B^i_j(h(\sx)))\right] 
& =\Pr[\mathfrak{E}] \E[g^{(-1)}(B^i_j(h(\sx)))\,|\,\mathfrak{E}]+\Pr[\mathfrak{E}^c] \E[g^{(-1)}(B^i_j(h(\sx)))\,|\,\mathfrak{E}^c]\notag\\
& \leq\Pr[\mathfrak{E}]
\E[g^{(-1)}(B^i_j(h(\sx)))\,|\,\mathfrak{E}]+\E[g^{(-1)}(B^i_j(h(\sx)))\,|\,\mathfrak{E}^c].
\label{eq8}
\end{align}
For the first term in \eqref{eq8}, we consider the following using \eqref{eq6}:
\begin{align}
\Pr[\mathfrak{E}] \E[g^{(-1)}(B^i_j(h(\sx)))~|~\mathfrak{E}]
&\leq\Big(1-\frac{1}{\cW_{\max}}\Big)^{h(\sx)-1}\cdot g^{(-1)}(B^i_j+h(\sx))\notag\\
&\leq\Big(1-\frac{1}{\sqrt{C(\sx)/2}}\Big)^{C(\sx)^{n}-1}\cdot g^{(-1)}\big(C(\sx)+C^{n}(\sx)\big)\notag\\
& =O(1),\label{eq9}
\end{align}
where the last inequality follows for $C(\sx)$ large enough, i.e.\
$\bW_{\max}$ large enough. In above we have used the definition
$h(\sx) = C(\sx)^n$ with $C(\sx) = \max\{g(\bA_{\max}), B_{\max}\}$
which is at least $\bW_{\max}^3$.  For the second term in
\eqref{eq8}, we consider the following using \eqref{eq7}:
\begin{eqnarray}
\E[g^{(-1)}(B^i_j(h(\sx)))~|~\mathfrak{E}^c]
&\leq&\sum_{k=1}^{\infty} g^{(-1)}(k)\cdot \Big(1-\frac{1}{\cW_{\max}}\Big)^{k-1}\notag\\
&\stackrel{(a)}{=}&O\left(g^{(-1)}(\cW_{\max}^2)\right)\notag\\
&\stackrel{(b)}{=}&O\left(g^{(-1)}\left(C(\sx)/2\right)\right),\label{eq10}
\end{eqnarray}
where (b) is from \eqref{eq:wmax} and for (a) we prove the following
technical proposition whose proof is presented in Appendix
\ref{sec:pfgeosum}.
\begin{proposition}\label{clm:geosum}
For $p\in (0,1)$,
$$\sum_{k=1}^{\infty} g^{(-1)}(k) \cdot (1-p)^{k}~=~O\left(
g^{(-1)}\left(p^{-2}\right)\right).$$
\end{proposition}
Combining \eqref{eq8}, \eqref{eq9} and \eqref{eq10}, the desired
conclusion of Lemma \ref{lem2-1} follows. This completes the proof
of Lemma \ref{lem2-1}.


\section{Proof of Lemma \ref{lem:PR}: $C(\sx) <
\bW_{\max}^3$}\label{sec:proof.b}

We remind that the goal is to establish that starting with state
$X(0) = \sx = (\bQ,\sig,\ba,\bA,\bB)$ such that $L(\sx)$ is large
enough and $C(\sx) = \max\{g(\bA_{\max}), \bB_{\max}\}<
\bW_{\max}^3$, after time $$h(\sx) = \frac{1}{2}
\exp\big(\exp(\log^{1/2} \bW_{\max})\big),$$ the expected value of
$L$ decreases by $$k(\sx) = \frac{\log^{1/2} \bW_{\max}}{2}
\exp\big(\exp(\log^{1/2} \bW_{\max})\big) ~= \log^{1/2} \bW_{\max} ~ h(\sx).$$ In other words,
\begin{align}\label{eq:lem1-1}
\E\big[L(h(\sx)) - L(0) \,|\, X(0) = \sx\big] & \leq - k(\sx).
\end{align}
We note that for proving Lemma \ref{lem:PR}, these will be the
definition of functions $h$ and $k$ (cf.\ \eqref{eq:def.h} and
\eqref{eq:def.k}) it concerns the case $C(\sx) < \bW_{\max}^3$. To
simplify notations, we will use notation $\E[\cdot]$ and
$\Pr[\cdot]$ instead of $\E[\cdot\,|\,X(0) = \sx]$ and
$\Pr[\cdot\,|\, X(0)=\sx]$ whenever clear from the context.

Similar to the proof of Lemma \ref{lem:PR} for the case $C(\sx)
\geq \bW_{\max}^3$ (presented in Section \ref{sec:proof.a}), we
start by obtaining some bound for $\bW_{\max}(\tau)$.
 To this end, note that
if $L(\sx)$ is large enough, then either $\bQ_{\max}$, $\bA_{\max}$
or $\bB_{\max}$ are large. Since $\bQ_{\max}$, $\bA_{\max}$ and
$\bB_{\max}$ are bounded in terms of $\bW_{\max}$, $\bW_{\max}$ is
necessarily large if $L(\sx)$ is large enough. Now for large enough
$L(\sx)$ and hence large enough $\bW_{\max}$, we have
\begin{eqnarray*}
\bW_{\max} &=& \max\left\{\log \bQ_{\max}, \exp(\sqrt{\log g(\bA_{\max})})\right\}\\
&\stackrel{(a)}{<}& \max\left\{\log \bQ_{\max}, \exp(\sqrt{3\log \bW_{\max}}\right\}\\
&\stackrel{(b)}{=}& \log \bQ_{\max},
\end{eqnarray*}
where (a) is from the condition $g(\bA_{\max})\leq C(\sx)<
\bW_{\max}^3$ and (b) is because $\bW_{\max}> \exp({\sqrt{3\log
\bW_{\max}}})$ for large enough $\bW_{\max}$. Henceforth, we shall assume
that
\begin{equation}\label{eq:wmaxlogq}
\bW_{\max} ~=~ \log \bQ_{\max}
\end{equation}
and consequently $\bQ_{\max}$ can be also assumed to be large enough
if $L(\sx)$ is large. Using this, we obtain the following lower
bound of $\bW_{\max}(\tau)$: for $\tau\leq h(\sx)$,
\begin{eqnarray}
\bW_{\max}(\tau)&\geq& \log \bQ_{\max}(\tau)~\stackrel{(a)}{\geq} ~
\log (\bQ_{\max}-h(\sx))\notag\\
&\stackrel{(b)}{=}&\log (\bQ_{\max}-o(\bQ_{\max}))\notag\\
&\stackrel{(c)}\geq& \frac12 \log
\bQ_{\max}~:=~\cW_{\min},\label{eq:wmin}
\end{eqnarray}
where (a) is from 1-Lipschitz property of $\bQ_{\max}(\cdot)$, (c)
holds for large enough $\bQ_{\max}$ and (b) is due to
$$h(\sx)=\frac{1}{2}
\exp\big(\exp(\log^{1/2} \bW_{\max})\big)=\frac{1}{2}
\exp\big(\exp(\log\log^{1/2} \bQ_{\max})\big)=o(\bQ_{\max}).$$ On
the other hand, $\bW_{\max}(\tau)$ can be upper bounded as follows:
for $\tau\leq h(\sx)$,
\begin{eqnarray}
\bW_{\max}(\tau)&\leq& \max\left\{\log \bQ_{\max}(\tau), \exp({\sqrt{\log g(\bA_{\max}(\tau))}})\right\}\notag\\
&\stackrel{(d)}{\leq}& \max\left\{\log (\bQ_{\max}+h(\sx)), \exp({\sqrt{\log g(\bA_{\max}+h(\sx))}})\right\}\notag\\
&\stackrel{(e)}{\leq} &\max\left\{\log (\bQ_{\max}+o(\bQ_{\max})), \log \bQ_{\max}\right\}\notag\\
&\leq& 2\cdot \log \bQ_{\max},\label{eq:wmax-1}
\end{eqnarray}
where (d) is from 1-Lipschitz properties of $\bQ_{\max}(\cdot)$,
$\bA_{\max}(\cdot)$ and (e) follows from below using $\bA_{\max} \leq g^{(-1)}(C(\sx))$, $C(\sx) < \bW_{\max}$:
\begin{eqnarray*}
\sqrt{\log g(\bA_{\max}+h(\sx))}&\leq&
\sqrt{\log g(g^{(-1)}(C(\sx))+h(\sx))}\\
&\leq&\sqrt{\log g\left(g^{(-1)}\left(\bW_{\max}^3\right)+h(\sx)\right)}\\
&=&\sqrt{\log g\left(g^{(-1)}\left(\log^3 \bQ_{\max}\right)+h(\sx)\right)}\\
&\stackrel{(f)}{\leq}&\sqrt{\log g\left(2\cdot h(\sx)\right)}\\
&=&\sqrt{\log g\left(\exp({\exp({\log\log^{1/2} \bQ_{\max}})})\right)}\\
&=&\log\log \bQ_{\max}.
\end{eqnarray*}
In above, for (f) one can check $$g^{(-1)}\left(\log^3
\bQ_{\max}\right)~\leq~ h(\sx)~=~ \frac12\exp({\exp({\log\log^{1/2}
\bQ_{\max}})})$$ for large enough $\bQ_{\max}$. Combining
\eqref{eq:wmin} and \eqref{eq:wmax-1}, it follows that for $\tau\leq
h(\sx)$,
\begin{equation}
\cW_{\min}~\leq~ \bW_{\max}(\tau)~\leq \cW_{\max},\label{eq:wminmax}
\end{equation}
where $\cW_{\min}:=\frac12\log \bQ_{\max}$ and $\cW_{\max}:=2 \log
\bQ_{\max}$.

\paraG{Three Lemmas}
Now we state the following key lemmas that will lead to
\eqref{eq:lem1-1}. We shall present their proofs in Section
\ref{sec:pflemqh2}, \ref{sec:pflemah2} and \ref{sec:pflembh2},
respectively.
\begin{lemma}\label{lem:qh2}
Given initial state $X(0)=\sx=(\bQ,\sig,\ba,\bA,\bB)$, let $C(\sx) <
\bW_{\max}^3$, $\lamb\in \Lamb$ and $\bQ_{\max}$ be large enough.
Then
$$\E\,\Big[\sum_i F(Q_i(h(\sx)))\Big]\leq \sum_i F(Q_i) - \Omega(\log\log \bQ_{\max})\cdot h(\sx).$$
\end{lemma}

\begin{lemma}\label{lem:ah2}
Given initial state $X(0)=\sx=(\bQ,\sig,\ba,\bA,\bB)$, let $C(\sx) <
\bW_{\max}^3$ and $\bQ_{\max}$ be large enough. Then for any $i$ and
$j\in \mathcal N(i)$
$$\E\,[A^i_j(h(\sx))^2]~=~ O(h(\sx)).$$
\end{lemma}

\begin{lemma}\label{lem:bh2}
Given initial state $X(0)=\sx=(\bQ,\sig,\ba,\bA,\bB)$, let $C(\sx) <
\bW_{\max}^3$ and $\bQ_{\max}$ be large enough. Then for any $i$ and
$j\in \mathcal N(i)$
$$\E\,[g^{(-1)}(B^i_j(h(\sx)))]~=~ O\left(g^{(-1)}(4\log^2 \bQ_{\max})\right).$$
\end{lemma}

\medskip

\paraG{Concluding \eqref{eq:lem1-1} using Lemma \ref{lem:qh2}, \ref{lem:ah2} and \ref{lem:bh2}}
These lemmas lead to the desired conclusion \eqref{eq:lem1-1} as
follows:
\begin{eqnarray*}
&&\E\,[L(h(\sx))-L(0)]\\
&&\qquad= \E\,\Big[\sum_i F(Q_i(h(\sx)))-\sum_i F(Q_i)\Big]+
\E\,\Big[\sum_{i,j} A^i_j(h(\sx))^2 - (A^i_j)^2\Big]\\
&&\qquad\qquad+\E\,\Big[\sum_{i,j} g^{(-1)}(B^i_j(h(\sx))) - g^{(-1)}(B^i_j)\Big]\\
&&\qquad\leq- \Omega(\log\log \bQ_{\max})\cdot h(\sx)+O(h(\sx))+
O\left(g^{(-1)}(4\log^2 \bQ_{\max})\right)\\
&&\qquad\stackrel{(a)}{\leq}- \Omega(\log\log \bQ_{\max})\cdot h(\sx)\\
&&\qquad=- \Omega(\log \bW_{\max})\cdot h(\sx)\\
&&\qquad\stackrel{(b)}{\leq}- \log^{1/2} \bW_{\max}\cdot h(\sx),
\end{eqnarray*}
where (a) is because $g^{(-1)}(4\log^2 \bQ_{\max})=o(h(\sx))$ for
our choice of $h(\sx)=\frac12\,\expo{\expo{\log\log^{1/2}
\bQ_{\max}}}$ and (b) holds for large enough $\bW_{\max}$. This
completes the proof of Lemma \ref{lem:PR} for the case $C(\sx) <
\bW_{\max}^3$.

\subsection{Proof of Lemma \ref{lem:qh2}}\label{sec:pflemqh2}
We start by observing that
\begin{eqnarray}
&&\E\Big[\sum_i F(Q_i(h(\sx)))-\sum_i F(Q_i)\Big]\notag\\
&&\qquad=
\sum_{\tau=0}^{h(\sx)-1}\E\Big[\sum_i F(Q_i(\tau+1))-\sum_i F(Q_i(\tau))\Big]\notag\\
&&\qquad=\sum_{\tau=0}^{h(\sx)-1} \sum_i
\E\left[F(Q_i(\tau+1))-F(Q_i(\tau))\right]\notag\\
&&\qquad\stackrel{(a)}{\leq} \sum_{\tau=0}^{h(\sx)-1} \sum_i
\E\left[(Q_i(\tau+1)-Q_i(\tau))\cdot f(Q_i(\tau+1))\right]\notag\\
&&\qquad=\sum_{\tau=0}^{h(\sx)-1} \sum_i
\E\left[(Q_i(\tau+1)-Q_i(\tau))\cdot
f(Q_i(\tau))\right]+O(h(\sx)),\label{eq1:pf2}
\end{eqnarray}
where (a) is due to the convexity of $F$ and the last inequality is from 1-Lipschitz property of
$Q_i(\cdot)$. For each term in the summation of \eqref{eq1:pf2}, we
consider the following.
\begin{eqnarray}
&&\E\left[(Q_i(\tau+1)-Q_i(\tau))\cdot f(Q_i(\tau))\right]\notag\\
&&\qquad=
\E\left[\left(A_i(\tau)-\sigma_i(\tau) \bind_{\{Q_i(\tau) > 0\}}\right)\cdot f(Q_i(\tau))\right]\notag\\
&&\qquad\stackrel{(a)}{=}
\E\left[A_i(\tau) \cdot f(Q_i(\tau))\right]-\E\left[\sigma_i(\tau) \cdot f(Q_i(\tau))\right]\notag\\
&&\qquad\stackrel{(b)}{=} \E\left[\lambda_i \cdot
f(Q_i(\tau))\right]-\E\left[\sigma_i(\tau) \cdot
f(Q_i(\tau))\right],\label{eq2:pf2}
\end{eqnarray}
where for (a) we use  $\bind_{\{Q_i(\tau) > 0\}}\cdot
f(Q_i(\tau))=f(Q_i(\tau))$ since $f(0)=0$; for (b) we use the fact
that $A_i(\tau)$, $Q_i(\tau)$ are independent random variables and
$\E[A_i(\tau)]=\lambda_i$. Now from \eqref{eq1:pf2} and
\eqref{eq2:pf2}, it follows that
\begin{eqnarray*}
&&\E\Big[\sum_i F(Q_i(h(\sx)))-\sum_i F(Q_i)\Big]\notag\\
&&\qquad=\sum_{\tau=0}^{h(\sx)-1} \sum_i
\E\left[(Q_i(\tau+1)-Q_i(\tau))\cdot f(Q_i(\tau))\right]+O(h(\sx))\\
&&\qquad=\sum_{\tau=0}^{h(\sx)-1}
\E\Big[\sum_i\lambda_i \cdot f(Q_i(\tau))-\sum_i\sigma_i(\tau) \cdot f(Q_i(\tau))\Big]+O(h(\sx))\\
&&\qquad\leq\sum_{\tau=0}^{h(\sx)-1}
\E\big[(1-\varepsilon)\Big(\max_{\brho\in\cI(G)} \brho\cdot
f(\bQ(\tau))\Big)-\sig(\tau) \cdot
f(\bQ(\tau))\big]+O(h(\sx)),\end{eqnarray*} where the last equality
is from $\lamb=[\lambda_i]\in \Lamb\subset (1-\varepsilon)
\Conv(\cI(G))$ for some $\varepsilon>0$ and the convex hull
$\Conv(\cI(G))$ of $\cI(G)$. Hence, for the proof of Lemma
\ref{lem:qh2}, it is enough to prove that
\begin{equation}
\sum_{\tau=0}^{h(\sx)-1}\E\big[(1-\varepsilon)\Big(\max_{\brho\in\cI(G)}
\brho\cdot f(\bQ(\tau))\Big)- \sig(\tau) \cdot
f(\bQ(\tau))\big]=-\Omega(\log\log \bQ_{\max})\cdot
h(\sx).\label{eq3:pf2}
\end{equation}
Further, it suffices to prove that for some $R=o(h(\sx))$
\begin{equation}
\sum_{\tau=R}^{h(\sx)-1}\E\big[(1-\varepsilon)\Big(\max_{\brho\in\cI(G)}
\brho\cdot f(\bQ(\tau))\Big)-\sig(\tau) \cdot
f(\bQ(\tau))\big]=-\Omega(\log\log \bQ_{\max})\cdot
(h(\sx)-R),\label{eq4:pf2}
\end{equation}
since the remaining terms in \eqref{eq3:pf2},  other than those
present in \eqref{eq4:pf2}, are dominated by \eqref{eq4:pf2} as
follows:
\begin{eqnarray*}
&&\sum_{\tau=0}^{R-1}
\E\big[(1-\varepsilon)\Big(\max_{\brho\in\cI(G)}
\brho\cdot f(\bQ(\tau))\Big)-\sig(\tau) \cdot f(\bQ(\tau))\big]\\
&&\qquad\leq~ \sum_{\tau=0}^{R-1} \E\big[\max_{\brho\in\cI(G)}
\brho\cdot f(\bQ(\tau))\big]~\leq~ \sum_{\tau=0}^{R-1}
\E\left[n\cdot f(\bQ_{\max}(\tau))\right]\\
&&\qquad\leq~ \sum_{\tau=0}^{R-1} \E\left[n\cdot
f\left(\expo{\bW_{\max}(\tau)}\right)\right]~=~ O(R) \cdot
f\left(\expo{\cW_{\max}}\right)\\
&&\qquad= o(h(\sx))\cdot \log\log \bQ_{\max},
\end{eqnarray*}
where the last equality is from $f(x)=\log\log x$, $\cW_{\max}\leq
2\log \bQ_{\max}$ (cf.\ \eqref{eq:wminmax}) and $R=o(h(\sx))$.

Now we will proceed toward proving \eqref{eq4:pf2}. Equivalently, we
will find some $R=o(h(\sx))$ such that for all $\tau\in [R,
h(\sx)-1]$,
\begin{equation}
\E\big[(1-\varepsilon)\Big(\max_{\brho\in\cI(G)} \brho\cdot
f(\bQ(\tau))\Big)-\sig(\tau) \cdot
f(\bQ(\tau))\big]=-\Omega(\log\log \bQ_{\max}).\label{eq5:pf2}
\end{equation}

\paraG{Three sub-Lemmas} The proof of \eqref{eq5:pf2} will be
established as a consequence of following
three lemmas. 
Their proof are presented in Section \ref{sec:lema}, \ref{sec:lemb}
and \ref{sec:lemc}, respectively.

\vspace{0.1in} \noindent {\bf Lemma A$~$} Let $\bmu(\tau)$ denote
the distribution of $\{\sig(\tau),\ba(\tau)\}$ at time $\tau$. Then,
    \begin{eqnarray*}
 \left\|  \bmu(\tau) -\bdel_{\{\sig,\ba\}}\cdot
 P(0)^{\tau}\right\|_{TV}&\leq&
O\Big(\sum_{s=0}^{\tau-1}
\E\left[\|P(s)-P(0)\|_{\infty}\right]\Big),\label{eq-1:relmu}
\end{eqnarray*}
where $\bdel_{x}$ is the Dirac distribution of singleton support $x$
and $P(\tau)$ denotes the transition matrix for the Markov chain
corresponding to the schedule $\sig(\tau)$, as described in
Section \ref{sec:mcofint} as well as Section \ref{sec:pfsum}
which is function of node weights $\bW(\tau)$ determined by
$\bQ(\tau)$ and $\bA(\tau)$ as per \eqref{eq:weight}.

\vspace{0.1in} \noindent {\bf Lemma B$~$} Let event
$\mathcal{E}_{\tau}$ at time $\tau$ be as follows:
    \begin{eqnarray*}
\mathcal{E}_{\tau}&:=& \Big\{X(\tau):W_i(\tau)\geq
\expo{\log\log^{\eta} \bQ_{\max}}~~\mbox{and}\\
&&\qquad\qquad g(A^i_j(\tau))\leq \log^4 \bQ_{\max}~~\mbox{for
all}~i,j\in \mathcal N(i)\Big\},
\end{eqnarray*}
where $\eta:=1/4^n$. Then, there exists $R_B={\bf
polylog}(\bQ_{\max})$ such that for $\tau< h(\sx)$,
\begin{eqnarray*}
&&(1-\varepsilon)\Big(\max_{\brho\in\cI(G)} \brho\cdot f(\bQ(\tau))\Big)-\E\left[\sig(\tau + R_B) \cdot f(\bQ(\tau))~\big|~ X(\tau)\in \mathcal{E}_{\tau}\right]\notag\\
&&\qquad\qquad\qquad\qquad\qquad\qquad\qquad\qquad\qquad\qquad\leq -\frac{\varepsilon}4\cdot \log\log \bQ_{\max}.\notag\\
\end{eqnarray*}

\vspace{0.05in} \noindent {\bf Lemma C$~$} There exists
$R_C=o(h(\sx))$ such that $\mathcal{E}_{\tau}$ happens with high
probability for $\tau\in [R_C, h(\sx)]$ i.e.\
\begin{equation*}
\Pr[\mathcal{E}_{\tau}]=1-o(1),
\end{equation*}
where recall $o(1)$ means that $o(1)\to 0$ as $\bQ_{\max}\to
\infty$. The $o(1)$ bound is uniform over all $\tau$.

\paraG{Remarks for Lemma A, B and C}
Before we derive the desired inequality \eqref{eq5:pf2} using the
above lemmas, some remarks are their role in establishing
it are in order.  To start with, the Lemma A captures
the evolution of the distribution of schedules. It is
used crucially to establish
Lemma B. Lemma B implies that \eqref{eq5:pf2} holds at time
$\tau+R_B$ if $\mathcal{E}_{\tau}$ happens at time $\tau$ and $R_B$
is small enough to guarantee $f(\bQ(\tau))\approx f(\bQ(\tau+R_B))$.
Lemma C indeed suggests that such event $\mathcal{E}_{\tau}$ happens
with high probability. This will essentially lead to \eqref{eq5:pf2}.

\paraG{Concluding \eqref{eq5:pf2}}
We choose $R$ for \eqref{eq5:pf2} as
$$R=R_B+R_C.$$
It is easy to check $R=o(h(\sx))$ since $R_B={\bf
polylog}(\bQ_{\max})=o(h(\sx))$ and $R_C=o(h(\sx))$. For $\tau\in
[R,h(\sx)-1]$, we break the left hand side of \eqref{eq5:pf2} into two
parts as follows:
\begin{eqnarray}
&&\E\big[(1-\varepsilon)\Big(\max_{\brho\in\cI(G)} \brho\cdot f(\bQ(\tau))\Big)-\sig(\tau) \cdot f(\bQ(\tau))\big]\notag\\
&&\qquad=
\Pr[\mathcal{E}_{\tau-R_B}]\cdot\E\big[(1-\varepsilon)\Big(\max_{\brho\in\cI(G)}
\brho\cdot f(\bQ(\tau))\Big) -\sig(\tau) \cdot f(\bQ(\tau))~\big|~
\mathcal{E}_{\tau-R_B}\big]\notag\\
&&\qquad\quad+\Pr[\mathcal{E}_{\tau-R_B}^c]\cdot
\E\big[(1-\varepsilon)\Big(\max_{\brho\in\cI(G)} \brho\cdot
f(\bQ(\tau))\Big)-\sig(\tau) \cdot f(\bQ(\tau))~\big|~
\mathcal{E}_{\tau-R_B}^c\big].\label{eq1:partd}
\end{eqnarray}
For the first term in \eqref{eq1:partd}, we obtain
\begin{eqnarray}
&&\Pr[\mathcal{E}_{\tau-R_B}]\cdot\E\big[(1-\varepsilon)\Big(\max_{\brho\in\cI(G)}
\brho\cdot f(\bQ(\tau))\Big)-\sig(\tau) \cdot f(\bQ(\tau))~\big|~
\mathcal{E}_{\tau-R_B}\big]\notag\\
&&\qquad\stackrel{(a)}{=} (1-o(1))\cdot
\E\big[(1-\varepsilon)\Big(\max_{\brho\in\cI(G)} \brho\cdot
f(\bQ(\tau))\Big)-\sig(\tau) \cdot f(\bQ(\tau))~\big|~
\mathcal{E}_{\tau-R_B}\big]\notag\\
&&\qquad\stackrel{(b)}{\leq} (1-o(1))\cdot
\E\big[(1-\varepsilon)\Big(\max_{\brho\in\cI(G)} \brho\cdot
f(\bQ(\tau-R_B))\Big)\notag\\
&&\qquad\qquad\qquad\qquad\qquad\qquad-\sig(\tau) \cdot
f(\bQ(\tau-R_B))~\big|~
\mathcal{E}_{\tau-R_B}\big]-O(f(R_B))\notag\\
&&\qquad\stackrel{(c)}{\leq}
-(1-o(1))\cdot\frac{\varepsilon}4\cdot \log\log \bQ_{\max}-O(f(R_B))\notag\\
&&\qquad\stackrel{(d)}{\leq} -\frac{\varepsilon}8\cdot \log\log
\bQ_{\max},\label{eq2:partd}
\end{eqnarray}
where (a) and (c) are from Lemma C and Lemma B, respectively. For
(b), we use 1-Lipschitz property of $Q_i(\cdot)$ and
$|f(x)-f(y)|<f(|x-y|)+O(1)$ for $f(x)=\log\log x$. For (d), we use
$$f(R_B)=f({\bf polylog}(\bQ_{\max}))=o(f(\bQ_{\max}))=o(\log\log
\bQ_{\max}).$$

For the second term in \eqref{eq1:partd}, we observe that
\begin{eqnarray}
&&\Pr[\mathcal{E}_{\tau-R_B}^c]\cdot
\E\big[(1-\varepsilon)\Big(\max_{\brho\in\cI(G)} \brho\cdot
f(\bQ(\tau))\Big)-\sig(\tau) \cdot f(\bQ(\tau))~\big|~
\mathcal{E}_{\tau-R_B}^c\big]\notag\\
&&\qquad\stackrel{(e)}{\leq} o(1)\cdot \E\big[\Big(\max_{\brho\in\cI(G)}
\brho\cdot f(\bQ(\tau))\Big)~\big|~
\mathcal{E}_{\tau-R_B}^c\big]\notag\\
&&\qquad\leq o(1)\cdot \E\big[n\cdot f(Q_{\max}(\tau))~\big|~
\mathcal{E}_{\tau-R_B}^c\big]\notag\\
&&\qquad\stackrel{(f)}{=} o(1)\cdot O(f(\bQ_{\max})) ~=~ o(\log\log
\bQ_{\max}),\label{eq3:partd}
\end{eqnarray}
where (e) is from Lemma C and (f) is due to
\begin{eqnarray*}
f(\bQ_{\max}(\tau))&\leq& f(\bQ_{\max}+\tau)~\leq~
f(\bQ_{\max}+h(\sx))\\
&= &f(\bQ_{\max}+ o(\bQ_{\max}))~=~O(f(\bQ_{\max})).
\end{eqnarray*}

Finally, combining \eqref{eq1:partd}, \eqref{eq2:partd} and
\eqref{eq3:partd}, the desired \eqref{eq5:pf2} follows as
\begin{eqnarray*}
\E\Big[(1-\varepsilon)\Big(\max_{\brho\in\cI(G)} \brho\cdot
f(\bQ(\tau))\Big)-\sig(\tau) \cdot f(\bQ(\tau))\Big] &\leq&
-\frac{\varepsilon}{16}\cdot \log\log \bQ_{\max}.
\end{eqnarray*}
This completes the proof of Lemma \ref{lem:qh2}.

\subsubsection{Proof of Lemma A}\label{sec:lema}
Let $\bmu(\tau+1:\tau)$ be the distribution of
$\{\sig(\tau+1),\ba(\tau+1)\}$ at time $\tau+1$ given network state
$X(\tau)$ at time $\tau$. From the definition of $P(\tau)$, we have
\begin{eqnarray*}
\bmu(\tau+1:\tau)& = & \bdel_{\{\sig(\tau),\ba(\tau)\}} P(\tau),
\end{eqnarray*}
where recall that $P(\tau)$ is a function of the network state $X(\tau)$
since node weights
$\bW(\tau)$ is decided by $\bQ(\tau)$ and $\bA(\tau)$ as per
\eqref{eq:weight}. By taking expectations on both sides of the above
equation, we obtain {\begin{eqnarray*}
\bmu(\tau+1)
&=&\E\left[\bdel_{\{\sig(\tau),\ba(\tau)\}} P(\tau)\right],
\end{eqnarray*}
where the expectation is with respect to the distribution of $X(\tau)$.
Using the above relation, we have
\begin{eqnarray*}
\bmu(\tau+1)&=&\E\left[\bdel_{\{\sig(\tau),\ba(\tau)\}} P(\tau)\right]\\
&=&\E\left[\E\left[\bdel_{\{\sig(\tau),\ba(\tau)\}} P(\tau)\,\Big|\,\bQ(\tau), \bA(\tau)\right]\right]\\
&=&\E\left[\E\left[\bdel_{\{\sig(\tau),\ba(\tau)\}}\,\Big|\,\bQ(\tau), \bA(\tau)\right]\cdot P(\tau)\right]\\
&=&\E\left[\tilde{\bmu}(\tau)\cdot P(\tau)\right],
\end{eqnarray*}
where the expectation is with respect to the distribution of
$\{\bQ(\tau),\bA(\tau)\}$ and we have used notation
\begin{eqnarray*}
\tilde{\bmu}(\tau)=\tilde{\bmu}(\bQ(\tau),
\bA(\tau)):=\E\left[\bdel_{\{\sig(\tau),\ba(\tau)\}}\,\Big|\,\bQ(\tau),
\bA(\tau)\right].
\end{eqnarray*}
This leads to the following recursive relation between
$\bmu(\tau+1)$ and $\bmu(\tau)$.
\begin{eqnarray*}
\bmu(\tau+1)
&=&\E\left[\tilde{\bmu}(\tau)\cdot P(\tau)\right]\\
&=&\E\left[\tilde{\bmu}(\tau)\cdot P(0)\right]
+\E\left[\tilde{\bmu}(\tau)\cdot (P(\tau)-P(0))\right]\\
&=&\E\left[\tilde{\bmu}(\tau)\right]\cdot P(0)
+e(\tau)\\
&=&\bmu(\tau)\cdot P(0) +e(\tau),
\end{eqnarray*}
where we define
\begin{eqnarray*}
e(\tau):=\E\left[\tilde{\bmu}(\tau)\cdot (P(\tau)-P(0))\right].
\end{eqnarray*}
By recursive application of this relation, we obtain
\begin{eqnarray*}
\bmu(\tau) &=&\bmu(0)\cdot P(0)^{\tau} +\sum_{s=0}^{\tau-1}
e(\tau-1-s)\cdot P(0)^s\notag\\
&=&\bdel_{\{\sig,\ba\}}\cdot P(0)^{\tau} +\sum_{s=0}^{\tau-1}
e(s)\cdot P(0)^{\tau-1-s}.
\end{eqnarray*}
Now we obtain the desired conclusion of Lemma A from the above
inequality as follows:
\begin{eqnarray*}
\left\|\bmu(\tau)-\bdel_{\{\sig,\ba\}}\cdot P(0)^{\tau}\right\|_{TV}
&=&\left\|\sum_{s=0}^{\tau-1} e(s)\cdot
P(0)^{\tau-1-s}\right\|_{TV}\\
&\leq&\sum_{s=0}^{\tau-1} \left\|e(s)\cdot
P(0)^{\tau-1-s}\right\|_{TV}\\
&\leq&O\left(\sum_{s=0}^{\tau-1} \left\|e(s)\right\|_{TV}\right)\\
&\leq&O\left(\sum_{s=0}^{\tau-1}
\E\left[\|P(s)-P(0)\|_{\infty}\right]\right),
\end{eqnarray*}
where we have used the fact that $P(0)^{\tau-1-s}$ (resp.
$\tilde{\bmu}(\tau)$) is a transition matrix (resp. distribution
vector) of finite dimension, independent of initial state $\sx$.
This completes the proof of Lemma A.

\subsubsection{Proof of Lemma B}\label{sec:lemb}
Recall that $\tau$ is time such that even $\mathcal{E}_{\tau}$ holds,
where
\begin{eqnarray*}
\mathcal{E}_{\tau}&:=& \Big\{X(\tau):W_i(\tau)\geq
\expo{\log\log^{\eta} \bQ_{\max}}~~\mbox{and}\\
&&\qquad\qquad g(A^i_j(\tau))\leq \log^4 \bQ_{\max}~~\mbox{for all}~i,j\in \mathcal N(i)\Big\},
\end{eqnarray*}
with $\eta:=1/4^n$. We wish to show the existence of $R_B$ so that
$R_B={\bf polylog}(\bQ_{\max})$ and
\begin{eqnarray*}
&&(1-\varepsilon)\Big(\max_{\brho\in\cI(G)} \brho\cdot f(\bQ(\tau))\Big)-\E\left[\sig(\tau + R_B) \cdot f(\bQ(\tau))~\big|~ X(\tau)\in \mathcal{E}_{\tau}\right]
~\leq~ -\frac{\varepsilon}4\cdot \log\log \bQ_{\max}.\notag
\end{eqnarray*}
To that end, we shall show that the above property holds for
$$R_B~:=~T_{\mix}\left(1/{\bQ_{\max}},n,2\log \bQ_{\max}\right),$$
where $T_{\mix}$ is defined as per \eqref{eq:tmix} as part of the statement of
Lemma \ref{lem:mixing}. Clearly, from definition $R_B={\bf polylog}(\bQ_{\max})$.
Now given network state $X(\tau)\in \mathcal{E}_{\tau}$ at time $\tau<h(\sx)$,
we have
\begin{equation*}
\left\|\bdel_{\{\sig(\tau),\ba(\tau)\}}\cdot
P(\tau)^{R_B}-\bpi(\tau)\right\|_{TV}~\leq~\frac1{\bQ_{\max}}~=~o(1),
\end{equation*}
where we let $\bpi(\tau)$ denote the unique stationary distribution
of $P(\tau)$ and use Lemma \ref{lem:mixing} with $W_{\max}(\tau)\leq
\cW_{\max}= 2\log \bQ_{\max}$ (cf.\ \eqref{eq:wminmax}). The above
equality suggests the following: for distribution
$\bmu(\tau+R_B:\tau)$ of $\sig(\tau+R_B)$ given network state
$X(\tau)$,
\begin{eqnarray}
&&\left\|\bmu(\tau +R_B:\tau)-\bpi(\tau)\right\|_{TV}\notag\\
&&\qquad\leq \left\|\bmu(\tau
+R_B:\tau)-\bdel_{\{\sig(\tau),\ba(\tau)\}}\cdot
P(\tau)^{R_B}\right\|_{TV}\notag\\
&&\qquad\qquad\qquad\qquad\qquad\quad\qquad\qquad+
\left\|\bdel_{\{\sig(\tau),\ba(\tau)\}}\cdot
P(\tau)^{R_B}-\bpi(\tau)\right\|_{TV}\notag\\
&&\qquad=O\Bigg(\sum_{s=\tau}^{\tau+R_B-1}
\E\left[\|P(s)-P(\tau)\|_{\infty}\right]\Bigg)+o(1)\notag\\
&&\qquad\stackrel{(a)}{\leq}
O\Bigg(\sum_{s=\tau}^{\tau+R_B-1}\E\big[\max_{i} \left|W_i(s)-W_i(\tau) \right|\big]\Bigg)+o(1)\notag\\
&&\qquad\stackrel{(b)}{\leq}
\sum_{s=\tau}^{\tau+R_B-1}O\Bigg(\frac{(s-\tau)\cdot {2\expo{\log\log^{\eta}\bQ_{\max}}}}{g^{(-1)}\left(\expo{\log\log^{2\eta} \bQ_{\max}}\right)}\Bigg)+o(1)\notag\\
&&\qquad\stackrel{(c)}{\leq}\frac{{\bf polylog}(\bQ_{\max})\cdot {2\expo{\log\log^{\eta}\bQ_{\max}}}}{g^{(-1)}\left(\expo{\log\log^{2\eta} \bQ_{\max}}\right)}+o(1)\notag\\
&&\qquad\stackrel{(d)}{=}o(1)+o(1)~=~o(1),\label{eq3:partb}
\end{eqnarray}
where (a) is from Proposition \ref{prop:changep} that is stated below (proof
presented in Appendix \ref{sec:pfchangep}); (b), (c) and
(d) follow from the Corollary \ref{cor1:wbound} in Appendix \ref{sec:propofW};
by definition $R_B={\bf polylog}(\bQ_{\max})$, $W_i(\tau)\geq \expo{\log\log^{\eta}\bQ_{\max}}$
due to event $\mathcal{E}_{\tau}$ and
$$\frac{g^{(-1)}\left(\expo{\log\log^{2\eta} \bQ_{\max}}\right)}{\expo{\log\log^{\eta}\bQ_{\max}}}={\bf superpolylog}(\bQ_{\max}).$$
\begin{proposition}\label{prop:changep}
Given two weights $\bW^1=[W^1_i]$ and $\bW^2=[W^2_i]$, let $P^1$ and
$P^2$ be the Markov chains (i.e.\ their transition matrices) on
$\Omega$ we described in Section \ref{sec:mcofint} using node
weights $\bW^1$ and $\bW^2$, respectively. Then,
$$\left|P^1_{x x^{\prime}}-P^2_{x x^{\prime}}\right|~=~O\big(\max_i\left|W_i^1-W_i^2\right|\big),\qquad\mbox{for
all}~ x,x^{\prime}\in \Omega.$$
\end{proposition}


From the above inequality \eqref{eq3:partb}, it follows that
\begin{eqnarray}
&&\E\left[\sig(\tau +R_B) \cdot \log \bW(\tau)~\big|~ X(\tau)\in
\mathcal{E}_{\tau}\right]\notag\\
&&\qquad{\geq}\E\left[\sig_{\bpi(\tau)} \cdot \log \bW(\tau)~\big|~
X(\tau)\in \mathcal{E}_{\tau}\right]
-\left\|\bmu(\tau+R_B:\tau)-\bpi(\tau) \right\|_{TV}\Big(\max_{\brho\in\cI(G)} \brho\cdot \log \bW(\tau)\Big)\notag\\
&&\qquad\stackrel{(a)}{\geq}\left(1-\left\|\bmu(\tau+R_B:\tau)-\bpi(\tau) \right\|_{TV}\right)\Big(\max_{\brho\in\cI(G)} \brho\cdot \log \bW(\tau)\Big)-O(1)\notag\\
&&\qquad=(1-o(1))\Big(\max_{\brho\in\cI(G)} \brho\cdot \log
\bW(\tau)\Big)-O(1),\label{eq4:partb}
\end{eqnarray}
where $\sig_{\bpi(\tau)}\in\cI(G)$ is the random variable drawn by
$\bpi(\tau)$ and for (a) we use the following proposition and the
product-form characterization of $\bpi(\tau)$ in Lemma \ref{lem:pi} (proof can
be found in Appendix).
\begin{proposition}[Gibbs' Maximal Principle]\label{prop:goodpi}
Let $T: \Omega \to \R$ and let $\cM(\Omega)$ be space of all
distributions on $\Omega$. Define $F : \cM(\Omega) \to \R$ as
    $$F(\bmu) ~=~ \E[T(x_{\bmu})] + H_{ER}(\bmu),$$
    where $x_{\bmu}\in\Omega$ in the random variable drawn by $\bmu$ and $H_{ER}(\bmu)$ is the standard discrete entropy of $\bmu$.
    Then, $F$ is uniquely maximized by the distribution $\bnu$, where
    $$ \nu_x ~=~ \frac{1}{Z}~ \expo{T(x)},\qquad\mbox{for any}~~x \in \Omega,$$
where $Z$ is the normalization constant (or partition function).
    Further, with respect to $\bnu$, we have
    $$ \E[T(x_{\bnu})] ~\geq~ \max_{x \in \Omega} T(x) - \log |\Omega|. $$
    \end{proposition}

\paraG{Concluding Lemma B} We further bound
the difference between $f(Q_i(\tau))$ and $\log W_i(\tau)$ as
\begin{eqnarray}
|f(Q_i(\tau))-\log W_i(\tau)| &=&
\Big|f(Q_i(\tau))-\max\Big\{f(Q_i(\tau)), \max_{j\in\cN(i)} \sqrt{\log g(A^i_j(\tau))}\Big\}\Big|\notag\\
&\leq& \max_{j\in\cN(i)} \sqrt{\log g(A^i_j(\tau))}\notag\\
&\stackrel{(a)}{\leq}&\sqrt{\log \left(\log^4 \bQ_{\max}\right)}\notag\\
&=&o(\log\log
\bQ_{\max})\notag\\
&=&o(f(\bQ_{\max})),\label{eq5:partb}
\end{eqnarray}
where (a) is because $X(\tau)\in\mathcal{E}_{\tau}$. Hence, we have
\begin{eqnarray*}
&&\E\left[\sig(\tau+R_B) \cdot f(\bQ(\tau))~\big|~ X(\tau)\in \mathcal{E}_{\tau}\right]\notag\\
&&\qquad \stackrel{(b)}{=}\E\left[\sig(\tau+R_B) \cdot \log
\bW(\tau)~\big|~ X(\tau)\in \mathcal{E}_{\tau}\right]
-o(f(\bQ_{\max}))\notag\\
&&\qquad\stackrel{(c)}{\geq} (1-o(1))\Big(\max_{\brho\in\cI(G)} \brho\cdot \log \bW(\tau)\Big)-O(1)-o(f(\bQ_{\max}))\notag\\
&&\qquad\stackrel{(d)}{\geq} (1-o(1))\Big(\max_{\brho\in\cI(G)} \brho\cdot f(\bQ(\tau))\Big)-o(f(\bQ_{\max}))\\
&&\qquad\stackrel{(e)}{\geq} (1-o(1))\Big(\max_{\brho\in\cI(G)}
\brho\cdot f(\bQ(\tau))\Big),
\end{eqnarray*}
where (b), (d) are from \eqref{eq5:partb}, (c) is due to
\eqref{eq4:partb} and (e) follows from
\begin{eqnarray}
&&\max_{\brho\in\cI(G)} \brho\cdot f(\bQ(\tau))~\geq~
f(\bQ_{\max}(\tau))~\geq~ f(\bQ_{\max}-\tau)\notag\\
&&\qquad\geq f(\bQ_{\max}-h(\sx))~=~f(\bQ_{\max}-o(\bQ_{\max}))~=~
\frac{1}2f(\bQ_{\max}),\label{eq:largeq}
\end{eqnarray}
for large enough $\bQ_{\max}$. Finally, we derive the desired
conclusion of Lemma B as
\begin{eqnarray*}
&&(1-\varepsilon)\max_{\brho\in\cI(G)} \brho\cdot f(\bQ(\tau))-\E\left[\sig(\tau+R_B) \cdot f(\bQ(\tau))~\big|~ X(\tau)\in \mathcal{E}_{\tau}\right]\notag\\
&&\qquad
\leq-(\varepsilon-o(1))\Big(\max_{\brho\in\cI(G)} \brho\cdot f(\bQ(\tau))\Big)~\leq-\frac{\varepsilon}2\cdot \Big(\max_{\brho\in\cI(G)} \brho\cdot f(\bQ(\tau))\Big)\notag\\
&&\qquad \leq-\frac{\varepsilon}2\cdot
f(\bQ_{\max}(\tau))~\leq-\frac{\varepsilon}4\cdot f(\bQ_{\max}),
\end{eqnarray*}
where the last inequality is from \eqref{eq:largeq} i.e.\
$f(\bQ_{\max}(\tau))\geq  f(\bQ_{\max})/2$. This completes the proof
of Lemma B.

\subsubsection{Proof of Lemma C}\label{sec:lemc}

We wish to show that there exists $R_C = o(h(\sx))$ so that
for any $\tau\in [R_C,h(\sx)]$, the event ${\mathcal E}_\tau$ holds.
For this, it is sufficient to establish that for any $\tau \in [R_C, h(\sx)]$,
the following holds:
\begin{eqnarray}
&&\Pr\left[g(A^i_j(\tau))\leq \log^4
\bQ_{\max}\right]~=~1-o(1),\label{eq3:partc}
\end{eqnarray}
\begin{eqnarray}
&&\Pr\left[W_i(\tau)\geq \expo{\log\log^{\eta}
\bQ_{\max}}\right]~=~1-o(1),\label{eq4:partc}
\end{eqnarray}
with the $o(1)$ being uniform over the choice of $\tau \in [R_C, h(\sx)]$. We
introduce some notations first:
$$\mathcal{L}_1:=\expo{\log\log^{1/4} \bQ_{\max}}\quad\mbox{and}\quad
\mathcal{L}_k:=\expo{\log^{1/4}
\mathcal{L}_{k-1}}\quad\mbox{for}~k\geq 2$$
$$\mathcal{T}_k:=\sum_{l=1}^k g^{(-1)}(\mathcal{L}_l/20)\cdot \log^{n+3} \bQ_{\max}.$$
It can be checked inductively that
$$\mathcal{L}_k=\expo{\log\log^{1/4^k} \bQ_{\max}}\qquad\mbox{and}\qquad
\mathcal{L}_n=\expo{\log\log^{\eta} \bQ_{\max}},$$ where recall that
$\eta=1/4^n$. Next, we shall show that both \eqref{eq3:partc} and \eqref{eq4:partc}
hold for the definition of $R_C$ as follows:
$$R_C:=\mathcal{T}_{n-1}=\sum_{k=1}^{n-1} g^{(-1)}\left(\expo{\log\log^{1/4^k} \bQ_{\max}}/20\right)\cdot \log^{n+3} \bQ_{\max}.$$
Observe that as per this definition, $R_C=o(h(\sx))$. Now we establish that indeed
\eqref{eq3:partc} and \eqref{eq4:partc} hold.

\paraG{Proof of \eqref{eq3:partc}} Define $T_0=0< T_1<T_2<\dots$ so that $T_m$ is the
$m$th time when $A^i_j(\cdot)$ is updated i.e.\ $B^i_j(T_m-1)\geq 2$ and
$B^i_j(T_m)=0$.  Define $\widehat{m}$ as
$$\widehat{m}~:=~\inf\left\{m:B^i_j(T_m-1)\geq g(\gamma)~~\mbox{and}~~m>1\right\},$$
where $\gamma=g^{(-1)}(\log^4 \bQ_{\max})-2$. In addition, note that
\begin{eqnarray*}
g(A^i_j)&\leq&C(\sx)~\leq~\bW_{\max}^3~=~\log^3 \bQ_{\max}\\
&<&\log^4 \bQ_{\max}-3~=~g(\gamma +2)-3~<~g(\gamma-1)
\end{eqnarray*}
for large enough $\bQ_{\max}$ since for large values $|g\rq{}(\cdot)| \ll 1$.
Thus, $A^i_j < \gamma -1$.

Now if $A^i_j(\tau)\geq \gamma +2 =g^{(-1)}(\log^4 \bQ_{\max})$, one can check
that $T_{\widehat m} \leq \tau$ since there should be at least two
updates before time $\tau$ which make $A^i_j(\cdot)$ increase
beyond $x$. Otherwise, $A^i_j(\cdot)$ should remain less than
$\gamma +1$ under the algorithm until time $\tau$ since
$A^i_j< \gamma-1$ in the beginning. Therefore, we have
\begin{eqnarray*}
&&\Pr[g(A^i_j(\tau))\geq \log^4 \bQ_{\max}]
~=~\Pr[A^i_j(\tau)\geq g^{(-1)}(\log^4 \bQ_{\max})]\\
&&\qquad\leq~\Pr[T_{\widehat{m}}\leq \tau]
~\stackrel{(a)}{\leq}~\sum_{k=1}^{\tau}\Pr[T_{\widehat{m}}=k]\\
&&\qquad\stackrel{(b)}{\leq}~\sum_{k=1}^{\tau}\Pr\left[a_j(s)=1~\mbox{for}~s=k-2,\dots,k-\lceil g(x)\rceil-1\right]\\
&&\qquad\stackrel{(c)}{\leq}~\sum_{k=1}^{\tau}\left(1-\frac1{\cW_{\max}}\right)^{\lceil
g(x)\rceil-1} ~\leq~{\tau}\cdot \left(1-\frac1{2\log
\bQ_{\max}}\right)^{\lceil g(x)\rceil-1}\\
&&\qquad\stackrel{(d)}{\leq}~ O\left(\frac{\tau}{\bQ_{\max}}\right)
~\stackrel{(e)}{=}~ O\left(\frac{h(\sx)}{\bQ_{\max}}\right) ~=~o(1),
\end{eqnarray*}
where (a) is from the union bound; for (b) we utilize the fact
$\widehat{m}>1$; for (c) one can observe that under the algorithm
the probability that some node $j$ keeps attempting to transmit
consecutively (without stopping) for some time interval of length
$y$ is at most $\left(1-\frac1{\cW_{\max}}\right)^{y-1}$ ; (d) is
due to $x=g^{(-1)}(\log^4 \bQ_{\max})-2$; (e) is from
$h(\sx)=o(\bQ_{\max})$. This completes the proof of
\eqref{eq3:partc}.

\paraG{Proof of \eqref{eq4:partc}} We shall utilize the following
result crucially whose proof is presented in Appendix \ref{sec:pfcorkey}.
\begin{proposition}\label{cor:key}
Consider $i \in V$, $j\in \mathcal{N}(i)\union \{i\}$, $W>0$ and
network state $X(\tau)=\{\bQ(\tau),\sig(\tau),\ba(\tau),\bA(\tau),\bB(\tau)\}$
at time $\tau\leq h(\sx)$. Suppose that $\bQ_{\max}$ is large enough
and
\begin{equation}\label{eq1:clmkey}
W_i(\tau)~>~W~\geq~ \expo{\log\log^{\delta}
\bQ_{\max}}\qquad\mbox{for some }\delta>0.
\end{equation}
Then,
$$\Pr\left[W_j(\tau+ s)~\geq~ \expo{\log^{1/4}W}\right]\geq1-o(1),$$ where
$s:=g^{(-1)}\left(W/20\right)\cdot \log^{n+3} \bQ_{\max}$. Here $o(1)$ is uniform
over choice of any $\tau \leq h(\sx)$.
\end{proposition}

Let $i^* \in \arg\max_i Q_i(\tau)$. For any node $j$, one can construct
a path $j_1=i^*,j_2,\dots,j_n=j$ of length $n$ by allowing repetition. We
recall the definition of
$\mathcal{L}_k$ and $\mathcal{T}_k$.
$$\mathcal{L}_1=\expo{\log\log^{1/4} \bQ_{\max}}\qquad\mbox{and}\qquad
\mathcal{L}_k=\expo{\log^{1/4} \mathcal{L}_{k-1}}~~~\mbox{for}~k\geq
2$$
$$\mathcal{T}_0=0\qquad\mbox{and}\qquad
\mathcal{T}_k=\sum_{l=1}^k g^{(-1)}(\mathcal{L}_l/20)\cdot
\log^{n+3} \bQ_{\max}.$$
Also define
$\mathfrak{E}_k$ as
$$\mathfrak{E}_k~:=~\left\{X(\tau+\mathcal{T}_k):W_{j_{k+1}}(\tau+\mathcal{T}_k)\geq \mathcal{L}_{k+1}\right\}.$$
Now consider the following proposition:
\begin{proposition}\label{cor:key2}
For $k=1,\dots, n-1$ and $\tau\leq h(\sx)-\mathcal{T}_{n-1}$,
$$\Pr[\mathfrak{E}_k~|~\mathfrak{E}_1,\dots,\mathfrak{E}_{k-1}]~\geq~
1-o(1).$$
In above, $o(1)$ is uniform over choice of $\tau \leq h(\sx) - \mathcal{T}_{n-1}$.
\end{proposition}
\begin{proof}
We will prove Proposition \ref{cor:key2} by induction. The base case
$k=1$ follows from
\begin{eqnarray*}
W_{j_1}(\tau)&=&W_{i^*}(\tau)~\geq~ \log Q_{i^*}(\tau)~\geq~
\log (Q_{i^*}-\tau)\\
&\geq& \log (\bQ_{\max}-h(\sx))~\geq~ \log (\bQ_{\max}-o(\bQ_{\max}))\\
&\geq& \frac12 \log \bQ_{\max} ~\geq~ \expo{\log\log^{1/4}
\bQ_{\max}}\\
&=&\mathcal{L}_1,
\end{eqnarray*}
where inequalities hold for large enough $\bQ_{\max}$. It is easy to
establish the necessary induction step using Proposition \ref{cor:key} 
and $\mathcal{L}_k=\expo{\log\log^{\delta} \bQ_{\max}}$ with
$\delta=1/4^k$. This completes the proof of Proposition
\ref{cor:key2}.
\end{proof}
Therefore, Proposition \ref{cor:key2} implies that for $\tau\in [0,
h(\sx)-R_C]$,
\begin{eqnarray*}
\Pr\left[W_i(\tau+R_C)\geq \expo{\log\log^{\eta}
\bQ_{\max}}\right]&=&
\Pr\left[W_j(\tau+\mathcal{T}_{n-1})\geq \mathcal{L}_{n}\right]\\
&=&\Pr\left[W_{j_n}(\tau+\mathcal{T}_{n-1})\geq \mathcal{L}_{n}\right]\\
&=& \Pr\left[\mathfrak{E}_{n-1}\right]\\
 &=&\prod_{k=1}^{n-1}  \Pr\left[\mathfrak{E}_{k}~|~\mathfrak{E}_1,\dots,\mathfrak{E}_{k-1}\right]\\
 &\geq& (1-o(1))^{n-1}\\
  &\geq& 1-o(1).
\end{eqnarray*}
Note that in above, $n$ is a constant and $o(1)$ is with respect to scaling
of network state such as $\bQ_{\max}$. This completes the proof of \eqref{eq4:partc}.

\subsection{Proof of Lemma \ref{lem:ah2}}\label{sec:pflemah2}

We first state the following key proposition for the proof of Lemma \ref{lem:ah2}.

\begin{proposition}\label{clm:forlemah2}
If $C(\sx)\leq \bW_{\max}^3$, then
$$\Pr[A^i_j(\tau)> k]~=~\exp\Big(-\frac{g(k)}{\cW_{\max}}\Big)\cdot O(\tau),$$
for $\tau\leq h(\sx)$ and $k>g^{(-1)}\left(\bW_{\max}^3\right)$. Recall that
$\cW_{\max}$ is defined as per \eqref{eq:wminmax}.
\end{proposition}
\begin{proof}
First note that it is enough to consider the case when $k$ is an integer.
Let random time $T_0=0<T_1<T_2<\dots$ such that $T_m$ is the $m$th
time when $A^i_j(\cdot)$ is updated i.e.\ $B^i_j(T_m-1)\geq2$ and
$B^i_j(T_m)=0$. 
We define
$$\widehat{m}:=\inf\left\{m:B^i_j(T_m-1)\geq g(k-1)~~\mbox{and}~~m> 1\right\}.$$
Now observe that if $A^i_j(\tau)> k$,
then $$A^i_j(\tau)~>~ k~>~ A^i_j$$ since $A^i_j\leq
g^{(-1)}(C(\sx))\leq g^{(-1)}\left(\bW_{\max}^3\right)<k$.
Hence, if $A^i_j(\tau)> k$, $T_{\widehat m}\leq \tau$ since there should be at least two updates before time $\tau$
which make $A^i_j(\cdot)$ increase beyond  $k-1$. Otherwise,
$A^i_j(\cdot)$ should keep less than
$k+1$ under the algorithm until time $\tau$. Using this observation, we have
\begin{eqnarray*}
\Pr[A^i_j(\tau)\geq k]
&\leq&\Pr[T_{\widehat{m}}\leq \tau]\\
&\stackrel{(a)}{\leq}&\sum_{l=1}^{\tau}\Pr[T_{\widehat{m}}=l]\\
&\stackrel{(b)}{\leq}&\sum_{l=1}^{\tau}\Pr\left[a_j(s)=1~\mbox{for}~s=l-2,\dots,l-g(k-1)-1\right]\\
&\stackrel{(c)}{\leq}&\sum_{l=1}^{\tau}\Big(1-\frac1{\cW_{\max}}\Big)^{g(k-1)-1}\\
&\leq&{\tau}\cdot \Big(1-\frac1{\cW_{\max}}\Big)^{g(k-1)-1}\\
&=&\exp\Big(-\frac{g(k)}{\cW_{\max}}\Big)\cdot O(\tau),
\end{eqnarray*}
where (a) is from the union bound; (b) is from $\widehat{m}>1$; for
(c) one can observe that under the algorithm the probability that
some node $j$ keeps attempting to transmit consecutively (without
stopping) for some time interval of length $y$ is at most
$\left(1-\frac1{\cW_{\max}}\right)^{y-1}$. This completes the proof
of Proposition \ref{clm:forlemah2}.
\end{proof}
\paraG{Completing proof of Lemma \ref{lem:ah2}} We derive the following inequalities.
\begin{align*}
&\E\left[A^i_j(h(\sx))^2\right] ~=~\sum_{k=1}^{\infty} \Pr[A^i_j(h(\sx))= k]\cdot k^2\\
&\qquad=~\sum_{k=1}^{\scriptsize \sqrt{h(\sx)}-1} \Pr[A^i_j(h(\sx))=
k]\cdot k^2+
\sum_{k=\sqrt{h(\sx)}}^{\infty} \Pr[A^i_j(h(\sx))= k]\cdot k^2\\
&\qquad\leq~ h(\sx) + \sum_{k=\sqrt{h(\sx)}}^{\infty}
\Pr[A^i_j(h(\sx))= k]\cdot k^2\\&\qquad
\stackrel{(a)}{\leq}~ h(\sx) + \sum_{k=\sqrt{h(\sx)}}^{\infty} \frac1{k^4}\cdot O(h(\sx))\cdot k^2\\
&\qquad \leq~ h(\sx) + O(h(\sx))\cdot
\sum_{k=\sqrt{h(\sx)}}^{\infty} \frac1{k^2}~ =~O(h(\sx)),
\end{align*}
where for (a) we use the following inequality:
\begin{equation}
\Pr[A(h(\sx))= k]~\leq ~ \frac1{k^4}\cdot
O(h(\sx)),\qquad\mbox{for}~k\geq
\sqrt{h(\sx)}.\label{eq1:cor:expofa}
\end{equation}
Hence, it is enough to show \eqref{eq1:cor:expofa} to complete the
proof of Lemma \ref{lem:ah2}.

From Proposition \ref{clm:forlemah2}, it suffices to prove that
\begin{eqnarray*}
\exp\Big(-\frac{g(k)}{\cW_{\max}}\Big)&\leq&\frac1{k^4},\qquad\mbox{for}~k\geq
\sqrt{h(\sx)}.
\end{eqnarray*}
By taking the double-logarithm (i.e.\ $\log\log$) on both sides of the above inequality
and using $g(k)=\expo{\log\log^4 k }$, we
have the equivalent inequality as
\begin{eqnarray*}
\log\log^4 k -\log \cW_{\max}\geq \log 4+\log\log k.
\end{eqnarray*}
One can check the above inequality holds if $\log\log^4 k \geq 2\log
\cW_{\max}$ since $\cW_{\max}$ is large enough. Equivalently, the
desired condition for $k$ is
\begin{eqnarray*}
k &\geq& \expo{\expo{(2\log \cW_{\max})^{1/4}}}.
\end{eqnarray*}
Finally, $k\geq \sqrt{h(\sx)}$ satisfies the above condition since
\begin{eqnarray*}
\expo{\expo{(2\log \cW_{\max})^{1/4}}}& = &
\expo{\expo{\Theta\big(\log \log^{1/4} \bQ_{\max}\big)}}\\
&\leq& \frac1{\sqrt{2}}\cdot \expo{\frac12  \expo{\log\log^{1/2} \bQ_{\max}}}\\
&=&\sqrt{h(\sx)},
\end{eqnarray*}
where the first inequality is from the definition of $\cW_{\max}$ in \eqref{eq:wminmax} and the second inequality holds for large enough $\bQ_{\max}$.
This completes the proof of \eqref{eq1:cor:expofa}, hence the proof
of Lemma \ref{lem:ah2}.

\subsection{Proof of Lemma \ref{lem:bh2}}\label{sec:pflembh2}
To begin with, we note that the proof of Lemma \ref{lem:bh2} is
almost identical to that of Lemma \ref{lem2-1} in Section
\ref{sec:pflem2-1}. Let the random time
$\tau^*=\inf\{\tau:a_j(\tau)=0\}$ i.e.\ the first time when $j$ does
not attempt to transmit, and the event $\mathfrak{E}$ denotes
$\tau^*\geq h(\sx)$. Hence, if $\mathfrak{E}$ happens,
$B(h(\sx))=B+h(\sx)$ and
\begin{eqnarray}
\Pr[\mathfrak{E}]&\leq&\Pr\left[\mbox{$j$ attempts to transmit consecutively for time $\tau\in[0,h(\sx)-1]$}~\right]\notag\\
&\leq&\Big(1-\frac1{\cW_{\max}}\Big)^{h(\sx)-1},\label{eq1:pflembh2}
\end{eqnarray}
where the last inequality because $W_j(\tau)$ is uniformly bounded
above by $\cW_{\max}$.

On the other hand, if the event $\mathfrak{E}$ does not happen, $j$
stops its transmission before time $h(\sx)$, hence $B^i_j(\cdot)$
should set to $0$ before time $h(\sx)$. Based on this observation
and arguments similar to those used for establishing \eqref{eq1:pflembh2}, we
obtain
\begin{eqnarray}
\Pr[B^i_j(h(\sx))=k~|~\mathfrak{E}^c] &\leq&\Pr\big[\mbox{$j$
attempts to transmit consecutively}\notag\\
&&\qquad\qquad\quad \mbox{
for time $\tau\in [h(\sx)-k,h(\sx)-1]$}~\big]\notag\\
&\leq&\Big(1-\frac{1}{\cW_{\max}}\Big)^{k-1}.\label{eq2:pflembh2}
\end{eqnarray}
Further observe that
\begin{eqnarray}
\E\left[g^{(-1)}(B(h(\sx)))\right]&=&\Pr[\mathfrak{E}]\cdot \E[g^{(-1)}(B^i_j(h(\sx)))~|~\mathfrak{E}]
+\Pr[\mathfrak{E}^c]\cdot \E[g^{(-1)}(B^i_j(h(\sx)))~|~\mathfrak{E}^c]\notag\\
&\leq&\Pr[\mathfrak{E}]\cdot
\E[g^{(-1)}(B^i_j(h(\sx)))~|~\mathfrak{E}]
+\E[g^{(-1)}(B^i_j(h(\sx)))~|~\mathfrak{E}^c].
\label{eq3:pflembh2}
\end{eqnarray}
For the first term in \eqref{eq3:pflembh2}, we consider the
following using \eqref{eq1:pflembh2}.
\begin{eqnarray}
\Pr[\mathfrak{E}]\cdot \E[g^{(-1)}(B^i_j(h(\sx)))~|~\mathfrak{E}]&\leq&\Big(1-\frac{1}{\cW_{\max}}\Big)^{h(\sx)-1}\cdot g^{(-1)}(B^i_j+h(\sx))\notag\\
&\stackrel{(a)}{\leq}&\Big(1-\frac{1}{h(\sx)}\Big)^{h(\sx)-1}\cdot g^{(-1)}(h(\sx)+h(\sx))\notag\\
&\stackrel{(b)}{=}&o(1),\label{eq4:pflembh2}
\end{eqnarray}
where (a) follows from $\cW_{\max}\leq 2\log \bQ_{\max}\leq h(\sx)$
and $B^i_j\leq C(\sx)\leq \bW_{\max}^3=\log^3 \bQ_{\max}\leq h(\sx)$; one
can check (b) for large enough $h(\sx)$.

For the second term in \eqref{eq3:pflembh2}, we consider the
following using \eqref{eq2:pflembh2}.
\begin{eqnarray}
\E[g^{(-1)}(B^i_j(h(\sx)))~|~\mathfrak{E}^c]
&\leq&\sum_{k=1}^{\infty} g^{(-1)}(k)\cdot \Big(1-\frac{1}{\cW_{\max}}\Big)^{k-1}\notag\\
&\stackrel{(a)}{=}&O\left(g^{(-1)}(\cW_{\max}^2)\right)\notag\\
&\leq&O\left(g^{(-1)}\left(4\log^2
\bQ_{\max}\right)\right),\label{eq5:pflembh2}
\end{eqnarray}
where (a) is from Proposition \ref{clm:geosum}. Combining
\eqref{eq3:pflembh2}, \eqref{eq4:pflembh2} and \eqref{eq5:pflembh2}
completes the proof of Lemma \ref{lem:bh2}.

\section{Discussion}\label{sec:discussion}

As the main result of this paper, we presented a new medium-access
algorithm for an arbitrary wireless network where simultaneously transmitting
nodes must form an independent set of the network graph. The algorithm
is optimal in the sense that network Markov chain is positive-recurrent as long
as the imposed traffic demand can be satisfied by some scheduling algorithm. The
algorithm is entirely distributed: the only information it utilizes is its own
queue-size and the history of collision or successful transmission. In a sense,
this work settles an important question that has been of interest in distributed
computation, communication, probability and learning.

The algorithm we presented builds upon the work of \cite{RSS09} where the
algorithm required a bit of information exchange between neighbors per unit time.
Specifically, the key technical contribution of our work is to get rid of this requirement
by means of designing a novel learning mechanism that essentially estimates the
rate of a Bernoulli process with time varying rates. This learning mechanism could
be of much broader interest.


\pagebreak

\appendix

\section{Properties of Markov Chain in Section \ref{sec:mcofint}}

\subsection{Proof of Lemma \ref{lem:pi}}\label{sec:pflempi}

Consider the Markov chain $P$ with weights $\bW \in \Rp^n$ such that
$\bW_{\min}\geq 1$. Starting from $(\bzero,\bzero)$, from the
description of the Markov chain, it follows inductively that Markov
chain is always in state $(\sig,\ba) \in \Omega$ so that $\sig \leq
\ba$ component-wise, i.e.\ for any $i$, $\sig_i = 1 ~\Rightarrow a_i
= 1$. Further, transition from any such state $\bx = (\sig,\ba)
\in\Omega$ to $\bx^\prime =(\sig^\prime,\ba^\prime) \in \Omega$ is
possible if and only if (a) $\sig \cup \sig^\prime \in \cI(G)$, and
(b) for any $i$, $a_i = 1 ~\Rightarrow~a^\prime_j = 0, ~\forall ~j
\in \cN(i)$. From (a) and (b), it immediately follows (a') $\sig
\cup \sig^\prime \in \cI(G)$, and (b') for any $i'$, $a^\prime_{i'}
= 1~\Rightarrow~ a_{j'} = 0, ~\forall~ j' \in \cN(i')$. That is, if
transition from $\bx$ to $\bx^\prime$ is feasible,  then transition
from $\bx^\prime$ to $\bx$ is feasible. As per this, it immediately
follows that the state $(\bzero,\bzero)$ is reachable from (and to)
all feasible states. That is, starting from $(\bzero,\bzero)$ the
Markov chain $P$ is recurrent and let $\Omega^\prime$ be the
recurrent class. Further, there is a strictly positive probability
of being at state $(\bzero,\bzero)$. Therefore, Markov chain is
aperiodic on $\Omega^\prime$. More generally, it can be checked that
for any two states $\bx, ~\bx^\prime$ with positive $P_{\bx
\bx^\prime}$, it is equal to
\begin{equation}
P_{\bx \bx^{\prime}}~=~ c(\bx,\bx^{\prime})\cdot
\prod_{i\in \sig\setminus\sig^{\prime}} \frac1{W_i}
\cdot\prod_{i\in \sig \cap \sig^{\prime}}
\Big(1-\frac1{W_i}\Big),\label{eq:p}
\end{equation}
where $c(\bx, \bx^{\prime}) = 2^{ -
|\{i~:~a_i=0;~a^{\prime}_i=1\}|}$.


Now to establish \eqref{eq:statofP}, consider another Markov chain
$Q$ on $\Omega^\prime$ with $Q_{\bx\bx^\prime} > 0$ if and only if
$P_{\bx\bx^\prime} > 0$. Specifically, for $\bx = (\sig,\ba),
~\bx^\prime=(\sig^\prime,\ba^\prime) \in \Omega^\prime$ with $\bx
\neq \bx^\prime$ and $P_{\bx\bx^\prime} > 0$,
\begin{align}\label{eq:Q.transition}
Q_{\bx\bx^\prime} & = \frac1{2^n}\cdot  \prod_{i\in
\sig\setminus\sig^{\prime}} \frac1{W_i}\cdot \prod_{i\in \sig\cap
\sig^{\prime}} \Big(1-\frac1{W_i}\Big).
\end{align}
We choose the other `self-transitions' in $Q$ so as to make it a
valid probability transition matrix. This is indeed possible since
$Q_{\bx\bx^\prime} \leq P_{\bx\bx^\prime}$ for all $\bx \neq
\bx^\prime$ from \eqref{eq:p}. By definition $Q$ is recurrent and
apreiodic since $P$ is. Therefore, it has unique stationary
distribution, say $\qpi$. We claim that for any $\bx = (\sig,\ba)
\in \Omega^\prime$,
\begin{align}
\qpi_{(\sig,\ba)} & \propto \prod_{i\in \sig} W_i ~=~\expo{\sig\cdot \log \bW} 
\stackrel{\triangle}{=} \bW(\sig).\label{eq:qpi}
\end{align}
This is because, the following detailed-balanced condition is
satisfied by $Q, ~\qpi$: for any feasible transitions
$\bx=(\sig,\ba), ~\bx^\prime=(\sig^\prime,\ba^\prime) \in
\Omega^\prime$ with $\bx \neq \bx^\prime$,
\begin{align*}
\frac{Q_{\bx\bx^\prime}}{Q_{\bx^\prime\bx}} & = \frac{\prod_{i \in\sig\backslash \sig^\prime} \frac{1}{W_i} }{\prod_{i \in\sig^\prime\backslash \sig} \frac{1}{W_i}}  ~=~ \frac{\prod_{i \in\sig^\prime \backslash \sig} W_i }{\prod_{i \in\sig\backslash \sig^\prime} W_i} 
= \frac{\prod_{i \in \sig^\prime} W_i}{\prod_{i\in \sig} W_i} ~=~
\frac{\qpi_{\bx^\prime}}{\qpi_{\bx}}.
\end{align*}

Given characterization $\qpi$ and similarity between $Q$ and $P$, we
shall approximately characterize $\bpi$, the stationary distribution
of $P$, in form of $\qpi$. For this, we shall use the following
proposition.

\begin{proposition}\label{prop:compare}
Given a finite state space $\Sigma$, denoted by $\{1,\dots, N\}$,
consider two irreducible and apriodic Markov chains on $\Sigma$ with
transition probability matrices $A$ and $B$.  Let $A_{ij} > 0$ if
and only if $B_{ij} > 0$ for all $i, j \in \Sigma$. Define
\begin{align}
R(A,B) & = \max_{(i,j) : A_{ij} > 0} \Big(\frac{A_{ij}}{B_{ij}},
\frac{B_{ij}}{A_{ij}}\Big).
\end{align}
Let $\bpi^A$ and $\bpi^B$ be stationary distributions of $A$ and
$B$. Then,
\begin{align}
R(A,B)^{-N}  ~\leq \min_{i} ~\Big(\frac{\pi_i^A}{\pi_i^B}\Big) &
\leq ~\max_{i} ~\Big(\frac{\pi_i^A}{\pi_i^B}\Big) ~\leq R(A,B)^N.
\end{align}
Subsequently, the relative entropy between $\bpi^A$ and $\bpi^B$,
denoted by $D(\bpi^A | \bpi^B)$, is bounded above as $D(\bpi^A |
\bpi^B) \leq N \log R(A,B)$.
\end{proposition}

\noindent  Let $\bpi$ be the stationary distribution of $P$. For
each $\bx \in \Omega^\prime$, $\pi_\bx > 0$. Therefore, we can write
\begin{align}
\pi_\bx & \propto \exp\Big(F(\bx)\Big),
\end{align}
with $F: \Omega^\prime \to \R$ with $F((\bzero,\bzero)) = 0$. Now
from definition of $P$ and $Q$, it follows that
\begin{align*}
R(P, Q) & \leq 2^n.
\end{align*}
Using this, Proposition \ref{prop:compare} and the form of $\qpi$
(cf.\ \eqref{eq:qpi}), we have
\begin{align}
\exp\Big(F(\bx)\Big) & = \frac{\pi_{\bx}}{\pi_{(\bzero,\bzero)}}  \nonumber \\
 &  = \exp\Big(\sig \cdot \log W\Big) \frac{\pi_\bx}{\qpi_\bx} \frac{\qpi_{(\bzero,\bzero)}}{\pi_{(\bzero,\bzero)}} \nonumber \\
& = \exp\Big(\sig \cdot \log W+ U(\bx)\Big), \label{eq:c1}
\end{align}
where $| U(\bx) | \leq |\Omega^\prime| \log R(P,Q) \leq n 4^n  \log
2$. That is, we conclude that for any $\bx \in \Omega^\prime$
\begin{align}
\pi_\bx & \propto \exp\Big(F(\bx)\Big),
\end{align}
where $F(\bx) =\sig \cdot \log W + U(\bx)$ where $|U(\bx)| \leq n
4^n \log 2$. This completes the proof of Lemma \ref{lem:pi}.

\begin{proof}[Proposition \ref{prop:compare}]
{ The proof follows by use of characterization of stationary
distribution by means of Markov chain tree theorem (cf.
\cite{AT89}). Specifically, it characterizes the stationary
distribution of a finite state, irreducible and aperiodic Markov
chain, say $A$, as follows. Let $\cG = (\Sigma, \cE)$ be a directed
graph with $e = (i,j) \in \cE \subset \Sigma \times \Sigma$ if
$A_{ij} > 0$.  Then its stationary distribution, $\bpi^A$, is
characterized as
\begin{align}\label{eq:p-c-1}
\pi^A_i & \propto \sum_{T \in \cT(i)} \prod_{(k,\ell) \in T}
A_{k\ell},
\end{align}
where $\cT(i)$ is the set of all directed spanning trees of $\cG$
rooted at $i$; by $(k,\ell) \in T$ we mean directed edge $(k,\ell)$
that belongs to $T$.

As per hypothesis of Proposition, it follows that the transition
graph $\cG$ for Markov chains $A$ and $B$ are identical. Therefore,
from \eqref{eq:p-c-1} it follows that for any $i \in \Sigma$
\begin{align}\label{eq:p-c-2}
\frac{\pi^A_i}{\pi^B_i} & = \frac{\sum_{T \in \cT(i)}
\prod_{(k,\ell) \in T} A_{k\ell}}{\sum_{T \in \cT(i)}
\prod_{(k,\ell) \in T} B_{k\ell}}.
\end{align}
By definition of $R(A,B)$, the fact that number of edges in any tree
$T$ is $N = |\Sigma|$ and \eqref{eq:p-c-2} it follows that for any
$i \in \Sigma$
\begin{align}\label{eq:p-c-3}
R(A,B)^{-N} & \leq ~\frac{\pi^A_i}{\pi^B_i} ~\leq R(A,B)^N.
\end{align}
To establish bound on relative entropy of $\bpi^A, \bpi^B$, observe
that
\begin{align}
D(\bpi^A | \bpi^B) & = \sum_i \pi_i^A \log \frac{\pi^A_i}{\pi^B_i} \nonumber \\
                               & \leq \sum_i \pi_i^A \log R(A,B)^N \nonumber \\
                               & = \big(\sum_i \pi_i^A\big) N \log R(A,B)~=~ N \log R(A,B).
\end{align}
This completes the proof of Proposition \ref{prop:compare}.
}\end{proof}

\subsection{Proof of Lemma \ref{lem:mixing}}\label{sec:pflemmixing}

By definition of $\bpi_{\min}, ~\bpi_{\max}$ and from
\eqref{eq:statofP}, it follows that
\begin{align}
\bpi_{\min} & \geq \bpi_{\max} \exp\big(-n \bW_{\max} - n4^n \log 2\big) \nonumber \\
                   & \geq \frac{1}{|\Omega^\prime|} \exp\big(-n \bW_{\max} - n4^n \log 2\big) \nonumber \\
                   & \geq \exp\big(- n \bW_{\max} - n 4^n \log 4\big) ~=~ C_n \bW_{\max}^{-n}, \label{eq:pimin}
\end{align}
where we have used $|\Omega^\prime| \leq 4^n$, $n \log 2 \geq 1$ for
$n \geq 2$ and $C_n = \exp(-n4^n \log 4)$.

Let the time-reversal of $P$ be $P^*$, i.e.\ $P^*_{\bx\bx^\prime} =
\pi_{\bx^\prime} P_{\bx\bx^\prime}/\pi_{\bx}$ for any $\bx,
\bx^\prime \in \Omega^\prime$. It follows that $PP^*$ is a
reversible Markov chain on $\Omega^\prime$. Then $PP^*$ has real
eigenvalues taking values in $[-1,1]$: let they be $-1\leq
\lambda_{\min} \leq \dots \leq \lambda_2 \leq \lambda_1 \leq 1$.  It
can be checked that $PP^*$ is irreducible and aperiodic due to
structure of $P$. Therefore, it follows that $\lambda_1 = 1$,
$\lambda_{PP^*} = \max \{|\lambda_{\min}|, \lambda_2\} < 1$ and
$PP^*$ has the unique stationary distribution equals to $\bpi$, the
stationary distribution of $P$,  which corresponds to the
(normalized) eigenvector with eigenvalue $1$. In this setting, the
following is a well known \cite[Corollary 1.14]{MT06} bound on
`mixing time' of $P$: starting from any initial distribution $\bmu$
on $\Omega^\prime$,
\begin{align}\label{eq:mix.1}
\Big\|\frac{\bmu P^\tau}{\bpi} - 1 \Big\|_{2, \bpi} < \varepsilon, &
~\text{for} ~\tau \geq \frac{2}{1-\lambda_{PP^*}} \log
\frac{1}{\varepsilon \bpi_{\min}},
\end{align}
where the $\chi^2$ (chi-squared) distance between two distributions
on a finite state space (here $\Omega^\prime$) is defined as
\begin{align}
\Big\| \frac{\bnu}{\bmu} - 1 \Big\|_{2, \bmu} ~=~
\|\bnu-\bmu\|_{2,\frac{1}{\bmu}} ~=~ \sqrt{ \sum_{\bx \in
\Omega^\prime} \Big(\frac{\nu_\bx}{\mu_\bx}-1\Big)^2}.
\end{align}
Another distance of interest is total-variation, which is defined as
and related to $\chi^2$ distance as follows.
\begin{align}\label{eq:chisquared.tv}
\frac{1}{2} \Big\| \frac{\bnu}{\bmu} - 1 \Big\|_{2, \bmu} & \geq \|\bnu - \bmu\|_{TV} \nonumber \\
                                                             & = \sum_{\bx \in \Omega^\prime} | \nu_\bx - \mu_\bx |.
\end{align}
From \eqref{eq:mix.1} and \eqref{eq:chisquared.tv}, it follows that
\begin{align}\label{eq:mix.2}
\|\bmu P^\tau - \bpi \|_{TV} < \varepsilon, & ~\text{for} ~\tau \geq
\frac{2}{1-\lambda_{PP^*}} \log \frac{1}{2\varepsilon \bpi_{\min}}.
\end{align}
Thus, to bound the `mixing time' $T_{\mix}$ of $P$, 
we need an upper bound on $\lambda_{PP^*}$. To that end, we shall
bound the second largest eigenvalue $\lambda_2$ and the smallest
eigenvalue $\lambda_{\min}$ in that order.

For $\lambda_2$, by Cheeger's inequality \cite{DFK91, sinclair} it
is well known that
$$\lambda_{2} ~\le~ 1 - \frac{\Phi^2}{2}, $$
where $\Phi$ is the conductance of $PP^*$, defined as
\begin{align*}
 \Phi & =~ \min_{S \subset \Omega^\prime: \pi(S)\le\frac12} \frac{Q(S,S^c)}{\pi(S)\pi(S^c)},
\end{align*}
where $S^c = \Omega^\prime \backslash S$, $\pi(S)=\sum_{\bx\in
S}\pi_{\bx}$ and
\[
Q(S,S^c) = \sum_{\bx\in S,\by\in S^c}{\pi_\bx (PP^*)_{\bx\by}}.
\]
Therefore,
\begin{align}\label{eq:lem2}
    \Phi &\ge \min_{S\subset \Omega}{Q(S,S^c)}  ~ \ge  ~ \min_{(PP^*)_{\bx\by}\ne0} \pi_\bx (PP^*)_{\bx\by} \nonumber \\
    &\ge \bpi_{\min} ~ \Big(\min_{(PP^*)_{\bx\by}\ne0} (PP^*)_{\bx\by}\Big)  \nonumber \\
    & \geq \bpi_{\min} \Big(\min_{(PP^*)_{\bx\by}\ne0} P_{\bx \bzero} P_{\by \bzero} \frac{\pi_{\by}}{\pi_{\bzero}}\Big), \quad \text{where}~~\bzero=(\bzero,\bzero) \nonumber \\
    & \stackrel{(a)}{\geq}  C_n \bW_{\max}^{-n} \times (2\bW_{\max})^{-2n} \exp\big(-n4^n \ln 2\big) \nonumber \\
    &{\geq} C_n^{2} \bW_{\max}^{-3n},
\end{align}
where (a) is from \eqref{eq:pimin}, definition of $\bpi$ from
\eqref{eq:statofP} and definition of $C_n = 4^{-n4^n}$. Now for
$\lambda_{\min}$, we observe that for any $\bx \in \Omega^\prime$,
\begin{align*}
(PP^*)_{\bx\bx} & \geq P_{\bx \bzero}^2 \frac{\pi_\bx}{\pi_{\bzero}} \\
                             & \geq (2\bW_{\max})^{-2n} \exp(-n4^n \ln 2) \\
                              & \geq C_n^{2} \bW^{-2n}.
\end{align*}
Now it can be easily checked that $\lambda_{\min}\geq - 1 + 2
\min_{\bx} (P P^*)_{\bx\bx}$. From this and \eqref{eq:lem2}, it
follows that
\begin{align}
\lambda_{PP^*} & \leq 1 - \frac{1}{2} C_n^{-4} \bW_{\max}^{-6n}.
\label{eq:mixing.time}
\end{align}
Using \eqref{eq:pimin},\eqref{eq:mix.2}  and \eqref{eq:mixing.time},
it follows that starting from any initial distribution $\bmu$ on
$\Omega^\prime$, $\|\bmu P^\tau - \bpi\|_{TV} < \varepsilon$ for
\[
\tau \geq T_{\mix} ~\stackrel{\triangle}{=} 4 C_n^{-4}
\bW_{\max}^{6n} \log \Big(\frac{C_n^{-1}
\bW_{\max}^n}{2\varepsilon}\Big).
\]
This completes the proof of Lemma \ref{lem:mixing}.

\pagebreak

\section{Properties of $\bW(\cdot)$}\label{sec:propofW}

Here we establish deterministic properties of $\bW(\cdot)$ under the
algorithm that will be useful to prove the main theorem in this
paper. Specifically we establish that, for any $i$, $W_i(\cdot)$
changes slowly if it becomes large.
\begin{proposition}\label{clm:wbound}
There exists an absolute constant $w_o \geq 0$ so that for any node
$i$ and time $\tau$, if $W_i(\tau) \geq w_o$ then
\begin{align}\label{eq:wbound}
\big|W_i(\tau+1)-W_i(\tau)\big| & \leq
\frac{W_i(\tau)}{g^{(-1)}\left(\expo{\log^2 W_i(\tau)}\right)}.
\end{align}
\end{proposition}
\begin{proof}
We shall establish existence of large enough constant $w_o$ under
which the claimed result holds. To that end, given node $i$ and time
$\tau$, from definition of weight as per \eqref{eq:weight},
\begin{align*}
W_i(\tau) & = \exp\Big(\max\big \{ f(Q_i(\tau)), \sqrt{\max_{j\in
\mathcal{N}(i)} \log g(A^i_j(\tau))} \big\}\Big),
\end{align*}
where we use $f(x)=\log\log x$. Now $Q_i(\cdot)$ and $A_j^i(\cdot)$
for any $j \in \cN(i)$ changed by at most $\pm 1$ in unit time. That is,
they are uniformly 1-Lipschitz. Therefore,
\begin{align}
W_i(\tau+1) & \leq \exp\Big(\max\big\{f(Q_i(\tau)+1), \sqrt{\max_{j\in \mathcal{N}(i)} \log g(A^i_j(\tau)+1)}\big\}\Big), \label{eq1:clm2}\\
W_i(\tau+1) & \geq \exp\Big(\max\big\{f(Q_i(\tau)-1),
\sqrt{\max_{j\in \mathcal{N}(i)} \log g(A^i_j(\tau)-1)}\big\}\Big).
\label{eq2:clm2}
\end{align}
Using \eqref{eq1:clm2}, we shall establish an upper bound on
$W_i(\tau+1) - W_i(\tau)$. To that end, consider
$\exp(f(Q_i(\tau)+1))$: using Taylor's expansion
\begin{align}
\exp(f(Q_i(\tau)+1)) &\leq  \log (Q_i(\tau)+1) \notag\\
&\leq  \log Q_i(\tau) + \frac1{Q_i(\tau)}\notag\\
&\stackrel{(a)}{\leq} W_i(\tau) + \frac1{\exp(W_i(\tau))}
\label{eq3:clm2},
\end{align}
where (a) follows from the fact that $\log y+1/y  \geq \log x + 1/x$
if $0 < x \leq y$ for all $y$ large enough; $w_o$ is chosen so that
such is true when $y \geq w_o$; the fact that $\log Q_i(\tau)\leq
W_i(\tau)$ and the assumption that $W_i(\tau) \geq w_o$. In a
similar manner, using Taylor's expansion and the form of
the derivative of function $\exp(\sqrt{\log g(\cdot)})$, we have
\begin{align}
\exp\Big(\sqrt{\log g(A^i_j(\tau)+1)}\Big) 
& \leq  \exp\Big(\sqrt{\log g(A^i_j(\tau))}\Big) \left[ 1 + \frac{1}{A_j^i(\tau)}\frac{ 2 \log\log A_j^i(\tau)}{\log A_j^i(\tau)}\right] \nonumber \\
&\stackrel{(a)}{\leq} W_i(\tau) + W_i(\tau)
\frac{1}{g^{(-1)}\Big(\exp\big(\log^2 W_i(\tau)\big)\Big)}
\label{eq4:clm2}.
\end{align}
In above (a) follows because for all large enough $y$, $q(x) \leq
q(y)$ for $0 < x \leq y$ with $q(x) = \frac{1}{y} \frac{2\log\log
x}{\log x}$; for $y$ large enough $q(y) \leq \frac{1}{y}$; $w_o$ is
large enough so that these two inequalities are satisfied;
$W_i(\tau) \geq \exp(\sqrt{\log A_j^i(\tau)})$ and $W_i(\tau) \geq
w_o$. From \eqref{eq1:clm2}, \eqref{eq3:clm2} and \eqref{eq4:clm2},
it follows that
\begin{align*}
W_i(\tau+1)&\leq  W_i(\tau) + \max\Big\{\exp(-W_i(\tau)), W_i(\tau) \frac{1}{g^{(-1)}\Big(\exp\big(\log^2 W_i(\tau)\big)\Big)}\Big\} \\
& \leq W_i(\tau) +  W_i(\tau) \frac{1}{g^{(-1)}\Big(\exp\big(\log^2
W_i(\tau)\big)\Big)},
\end{align*}
where the last inequality follows from the fact that for $x$ large
enough, $\exp(-x) \leq x/g^{(-1)}(\exp(\log^2 x))$. This is because
$g^{(-1)}(y) = \exp(\exp(\log^{1/4} y))$,
\begin{align*}
\log^{1/4} \exp(\log^2 x) & = o(\log x), \quad \text{and hence} \\
g^{(-1)}(\exp(\log^2 x)) & = \exp(\exp(\log^{1/4} (\exp(\log^2 x))))
~=~ \exp(o(x)).
\end{align*}
We shall assume that $w_o$ is large enough to satisfy this and
$W_i(\tau) \geq w_o$.

This completes the proof of upper bound on $W_i(\tau+1)$ as desired
by Proposition of \ref{clm:wbound}. In the process, we implicitly
defined $w_o$: it is a constant large enough so that (i) $\log y+1/y
\geq \log x + 1/x$  if $0 < x \leq y$ for all $y \geq w_o$; (ii)
$q(x) \leq q(y) \leq 1/y$ for $0 < x \leq y$ with $q(x) =
\frac{1}{y} \frac{2\log\log x}{\log x}$ for all $y \geq w_o$; and
(iii) $\exp(-x) \leq x/g^{(-1)}(\exp(\log^2 x))$ for all $x \geq
w_o$.  In a similar manner (details are skipped here), an
appropriate  lower bound on $W_i(\tau+1)$ can be obtained (which
will lead to additional constraints on $w_o$). This completes the
proof of Proposition \ref{clm:wbound}.
\end{proof}
Following is an immediate corollary of Proposition \ref{clm:wbound}.
\begin{corollary}\label{cor1:wbound}
There exists a large enough constant $w_1$ so that for any node $i$,
if $W_i(0) \geq w_1$ then
\begin{align*}
\big|W_i(\tau)-W_i(0)\big| & \leq~
\frac{2W_i(0)}{g^{(-1)}\Big(\exp\big(\log^2 W_i(0)\big)\Big)} \tau,
\end{align*}
for $\tau \leq {g^{(-1)}\Big(\exp\big(\log^2
W_i(0)\big)\Big)}/{2W_i(0)}$.  Subsequently, for any $w_2 \geq w_1$,
\begin{align}
W_i(\tau) \leq w_2 + 1, & \quad \text{if} \quad W_i(0) \leq w_2 \label{eq:cor1.a}, \\
W_i(\tau) \geq w_2 - 1, & \quad \text{if} \quad W_i(0) \geq w_2
\label{eq:cor1.b},
\end{align}for $2\tau \leq {g^{(-1)}\Big(\exp\big(\log^2 w_2\big)\Big)}/{w_2}$.
\end{corollary}
\begin{proof}
We need to establish existence of large enough constant $w_1$ so
that claimed result holds. For this, we start by constraining $w_1
\geq w_o$, where $w_o$ is the constant from Proposition
\ref{clm:wbound}. In addition, we shall assume that $w_1$ is large
enough so that for all $y \geq w_1$, $g^{(-1)}(\exp(\log^2 x))/x
\leq g^{(-1)}(\exp(\log^2 y))/y$  as long as $0 < x \leq y$. Such is
a possibility since $g^{(-1)}(\exp(\log^2 x))$ scales much faster
than $x$. Further, function $g^{(-1)}(\exp(\log^2 x))/x =
\exp(o(x))$. Therefore, it can be shown that for large enough $x$,
\[
1 \ll \frac{g^{(-1)}(\exp(\log^2 x))}{2x} \leq
\frac{g^{(-1)}(\exp(\log^2 (x-1)))}{x-1}.
\]
We shall assume $w_1$ is chosen to be such large enough constant.
Now applying Proposition \ref{clm:wbound}, starting with $W_i(0)
\geq w_1$, it follows that
\begin{align*}
\big|W_i(1)-W_i(0) \big| & \leq~
\frac{W_i(0)}{g^{(-1)}\Big(\exp\big(\log^2 W_i(0)\big)\Big)}.
\end{align*}
As per the above bound and choice of $w_1$, $|W_i(1)-W_i(0)| \ll 1$.
By repeated application of Proposition \ref{clm:wbound} till the
summation of the right hand side of the above bound remains less
than $1$, we obtain
\begin{align*}
\big|W_i(\tau)-W_i(0) \big| & \leq \sum_{s=0}^{\tau-1}
\frac{W_i(s)}{g^{(-1)}\Big(\exp\big(\log^2 W_i(s)\big)\Big)}.
\end{align*}
Now for all such $s$, since $|W_i(s)-W_i(0)| \leq 1$, $W_i(0) \geq
w_1$ and above discussed properties of $w_1, ~g^{(-1)}$ we obtain
that for all such $\tau$
\begin{align*}
\big|W_i(\tau)-W_i(0) \big| & \leq \tau \frac{W_i(0)-1}{g^{(-1)}\Big(\exp\big(\log^2 (W_i(0)-1)\big)\Big)} \\
                                            & \leq \tau \frac{2W_i(0)}{g^{(-1)}\Big(\exp\big(\log^2 W_i(0)\big)\Big)}.
\end{align*}
Therefore, it follows that the above holds true for all $\tau$ such
that
\[
\tau \leq  \frac{g^{(-1)}\Big(\exp\big(\log^2
W_i(0)\big)\Big)}{2W_i(0)}.
\]
The remaining consequences \eqref{eq:cor1.a} and \eqref{eq:cor1.b}
follow immediately from this.  This complete the proof of Corollary
\ref{cor1:wbound}.
\end{proof}

\pagebreak

\section{Proofs of Auxiliary Results}

\subsection{Proof of Proposition
\ref{clm:geosum}}\label{sec:pfgeosum}
Using elementary calculus, it follows that
\begin{equation*}
\lim_{x\to\infty}\frac{g^{(-1)}(x)}{\expo{\sqrt{x}/8}}=0\qquad\mbox{and}\qquad
\lim_{x\to\infty}\frac{x g^{(-1)}(x^2/4)}{g^{(-1)}(x^2)}=0.
\end{equation*}
Hence, there exists a constant $C_1>0$ such that for $x>C_1$,
\begin{equation}
g^{(-1)}(x)\leq \expo{\sqrt{x}/8}\qquad\mbox{and}\qquad x\cdot
g^{(-1)}(x^2/4)\leq g^{(-1)}(x^2). \label{eq:prop1}
\end{equation}
If $k\geq p^{-2}/4$, it follows that
\begin{equation}
g^{(-1)}(k)~\leq~ \expo{\sqrt{k}/8} ~\leq~ \expo{p\cdot k/4} ~=~
\left(\expo{p/4}\right)^k ~\leq~ (1+p/2)^k, \label{eq:prop2}
\end{equation}
where the last inequality holds from $\expo{x}\leq 1+2x$ for $x\in
(0,1)$. Using this, 
\begin{eqnarray*}
&&\sum_{k=1}^{\infty} g^{(-1)}(k) \cdot (1-p)^k\\
&&\qquad=
\sum_{k=1}^{p^{-2}/4} g^{(-1)}(k) \cdot (1-p)^k +\sum_{k=p^{-2}/4+1}^{\infty} g^{(-1)}(k) \cdot (1-p)^k\\
&&\qquad\stackrel{(a)}{=} O\left(g^{(-1)}\left(p^{-2}/4\right)\cdot p^{-1}\right) +\sum_{k=p^{-2}/4+1}^{\infty} (1+p/2)^k \cdot (1-p)^k\\
&&\qquad = O\left(g^{(-1)}\left(p^{-2}/4\right)\cdot p^{-1}\right) +\sum_{k=p^{-2}/4+1}^{\infty}  (1-p/2)^k\\
&&\qquad = O\left(g^{(-1)}\left(p^{-2}/4\right)\cdot p^{-1}\right) +O\left(p^{-1}\right)\\
&&\qquad\stackrel{(b)}{=}
O\left(g^{(-1)}\left(p^{-2}\right)\right),
\end{eqnarray*}
where (a) is from \eqref{eq:prop2} and for (b) we use
\eqref{eq:prop1} under assuming $p^{-1}>C_1$. Otherwise, note that
(b) is trivial since $p^{-1}$ bounded above by constant $C_1$. This
completes the proof of Proposition \ref{clm:geosum}.

\subsection{Proof of Proposition
\ref{prop:changep}}\label{sec:pfchangep}
We recall the formula \eqref{eq:p}.
\begin{equation*}
P_{x x^{\prime}}~=~
c(x,x^{\prime})\cdot
\prod_{i\in \sig\setminus\sig^{\prime}} \frac1{W_i}
\cdot\prod_{i\in \sig \cap \sig^{\prime}}
\left(1-\frac1{W_i}\right),
\end{equation*}
where $c(x,x^{\prime})$ is some constant independent of $\bW=[W_i]$.
 Hence, we will consider $P_{x x^{\prime}}$
as a real-valued function in several variables $\{W_i\}$ i.e.\ $P_{x
x^{\prime}}=P_{x x^{\prime}}(\bW)$.

Now from the mean value theorem in several variables,
\begin{eqnarray*}
\left|P^1_{x x^{\prime}}-P^2_{x x^{\prime}}\right|&=& \left|\nabla
P_{x
x^{\prime}}(\cdot)\cdot (\bW^1-\bW^2)\right|\\
&\leq& \|\nabla P_{x x^{\prime}}(\cdot)\|_2\cdot \|\bW^1-\bW^2\|_2.
\end{eqnarray*}
Using this and \eqref{eq:p}, the desired conclusion follows since
one can easily check that $\|\nabla P_{x
x^{\prime}}(\cdot)\|_2=O(1)$ since each component of $\bW$ is always
at least $1$;
$\|\bW^1-\bW^2\|_2=O\left(\max_i\left|W_i^1-W_i^2\right|\right)$.
\subsection{Proof of Proposition
\ref{prop:goodpi}}\label{sec:pfgoodpi}

Observe that the definition of distribution $\bnu$ implies that for
any $x \in \Omega$,
$$T(x) = \log Z + \log \nu_{x}.$$ Using this, for any distribution
$\bmu$ on $\Omega$, we obtain
    \begin{align*}
      F(\bmu) &~=~ \sum_{x}\mu_{x}T(x) - \sum_{x}\mu_{x}\log \mu_{x}\\
      &~=~ \sum_{x}\mu_{x}(\log Z + \log \nu_{x}) - \sum_{x}{\mu_{x}\log\mu_{x}}\\
      &~=~ \sum_{x}{\mu_{x}\log Z} + \sum_{x}{\mu_{x}\log{\frac{\nu_{x}}{\mu_{x}}}}\\
      &~=~ \log Z + \sum_{x}{\mu_{x}\log{\frac{\nu_{x}}{\mu_{x}}}}\\
      &~\le~ \log Z + \log\biggl(\sum_{x}{\mu_{x}\frac{\nu_{x}}{\mu_{x}}}\biggr) \\
      &~=~ \log Z
    \end{align*}
    with equality if and only if $\bmu=\bnu$. To complete other claim of proposition,
    consider  $x^* \in \arg\max{T(x)}$. {Let
$\bmu$ be the Dirac distribution $\bdel_{x^*}$.} Then, for this
distribution
$$ F(\bmu) = T(x^*).$$
But, $F(\bnu) \geq F(\bmu)$. Also, the maximal entropy of any
distribution on $\Omega$ is $\log |\Omega|$. Therefore,
\begin{eqnarray}
T(x^*) & \leq & F(\bnu) \nonumber \\
         & = & \E[T(x_{\bnu})] + H_{ER}(\bnu) \nonumber \\
         & \leq &  \E[T(x_{\bnu})] + \log |\Omega|. \label{eq:ed10}
\end{eqnarray}
Re-arrangement of terms in \eqref{eq:ed10} will imply the second
claim of Proposition \ref{prop:goodpi}. This completes the proof of
Proposition \ref{prop:goodpi}.

\pagebreak

\section{Proof of Proposition \ref{cor:key}}\label{sec:pfcorkey}

Recall that the Proposition \ref{cor:key} assumes that $C(\sx) \leq \bW_{\max}^3$.
As per the statement of Proposition \ref{cor:key}, we wish to prove that
$$ \Pr\left[W_j(\tau+s )~\geq~ \expo{\log^{1/4}W}\right] \geq 1- o(1), $$
for choice of $ s$ such that
\begin{eqnarray*}
s &=&g^{(-1)}\left(W/20\right)\cdot \log^{n+3} \bQ_{\max}\\
&\stackrel{(a)}{\leq}&g^{(-1)}\left(W/20\right)\cdot \expo{(n+3)\log^{1/\delta} W}\\
&\stackrel{(b)}{\leq}& ~ \frac{g^{(-1)}\left(\expo{\log^2
W}\right)}{2W},
\end{eqnarray*}
where (a) is from the condition $W\geq \expo{\log\log^{\delta}
\bQ_{\max}}$ and one can check (b) for large enough $W$ (depending
on $\delta$).

Now for the case of $j = i$, from above and Corollary \ref{cor1:wbound}, we have
\begin{eqnarray*}
W_j(\tau+s)~\geq~W-1~\geq~ \expo{\log^{1/4}W},\qquad\mbox{with
probability}~1,
\end{eqnarray*}
where the last inequality holds for large enough $W$.

Now consider the case $j\neq i$. In this case, we have
\begin{eqnarray*}
&&\Pr\left[W_j(\tau+s)~\geq~
\expo{\log^{1/4}W}\right]\\
&&\qquad\stackrel{(a)}{\geq}
\Pr\left[\expo{\sqrt{\log g(A^j_i(\tau+s))}}~\geq~ \expo{\log^{1/4}W}\right]\\
&&\qquad=\Pr\left[g(A^j_i(\tau+s))~\geq~ \expo{\log^{1/2}W}\right]\\
&&\qquad\stackrel{(b)}{\geq}\Pr\left[g(A^j_i(\tau+s))~\geq~ W/20\right]\\
&&\qquad\stackrel{(c)}{\geq} 1-o(1),
\end{eqnarray*}
where (a) is from definition of $W_j$; for (b) we use
$\expo{\log^{1/2}W}<W/20$ for large enough $W$; (c) is due to the
following lemma. This completes the proof of Proposition
\ref{cor:key}.

\begin{lemma}\label{clm:key}
Consider given $i$, $j\in \mathcal{N}(i)$, $W>0$ and network state
$X(\tau)=\{\bQ(\tau),\sig(\tau),\ba(\tau), \bA(\tau),\bB(\tau)\}$ at
time $\tau\leq h(\sx)$. Suppose that $\bQ_{\max}$ is large enough and
\begin{equation}\label{eq1:clmkey}
W_i(\tau)~>~W~\geq~ \expo{\log\log^{\delta}
\bQ_{\max}}\qquad\mbox{for some }\delta>0.
\end{equation}
Then,$$\Pr\left[g(A^j_i(\tau+s))~\geq~ W/20\right]\geq
1-o(1),$$ where $s =g^{(-1)}\left(W/20\right)\cdot \log^{n+3}
\bQ_{\max}$.
\end{lemma}
\begin{proof} First consider the case when $g(A^j_i(\tau))> W/10$.
1-Lipschitz property of $A^j_i(\cdot)$ implies that
\begin{equation}
g(A^j_i(\tau+\tau^{\prime}))\geq W/20,\qquad\mbox{for all}~
\tau^{\prime} \leq
g^{(-1)}(W/10)-g^{(-1)}(W/20).\label{eq1:pfclmkey}
\end{equation}
On the other hand, we have
\begin{eqnarray}
\frac{g^{(-1)}(W/10)}{g^{(-1)}(W/20)}
&=&\expo{\expo{\log^{1/4} \frac{W}{10}}-\expo{\log^{1/4} \frac{W}{20}}}\notag\\
&\stackrel{(a)}{\geq}&
\expo{\expo{\log^{1/4} \frac{W}{10}}\cdot \frac14\log^{-3/4}\frac{W}{20}}\notag\\
&\stackrel{(b)}{=}&\mbox{\bf
superpolylog}(\bQ_{\max}).\label{eq2:pfclmkey}
\end{eqnarray}
where for (a) we use $d(x)-d(x/2)\geq d^{\prime}(x/2)\cdot x/2$ with
$d(x)=\expo{\log^{1/4} x}$ and $x=W/10$; (b) is due to $W\geq
\expo{\log\log^{\delta} \bQ_{\max}}$. Therefore, it follows that
$$g(A^j_i(\tau+s))~\geq~ W/20$$
since
\begin{equation}\label{eq3:pfclmkey}
s =g^{(-1)}\left(W/20\right)\cdot {\bf polylog}(\bQ_{\max})\ll
g^{(-1)}(W/10)-g^{(-1)}(W/20),
\end{equation}
where the inequality is from \eqref{eq1:pfclmkey},
\eqref{eq2:pfclmkey}, and large enough $\bQ_{\max}$.

Now consider the second case when $g(A^j_i(\tau))\leq W/10$. As the
first step, we will find some absolute upper and lower bounds of
$W_i(\tau+\tau^{\prime})$ and $g(A^j_i(\tau+\tau^{\prime}))$ for
$\tau^{\prime}\leq s$. Based on these bounds, we will
construct a martingale with respect to $g(A^j_i(\cdot))$ to control
$g(A^j_i(\tau+\Delta ))$, which is indeed similar to the strategy we
use for the proof of Lemma \ref{lem2} in Section \ref{sec:pflem2}.

\paraG{First step: Bounds for $W_i(\tau+\tau^{\prime})$, $g(A^j_i(\tau+\tau^{\prime}))$}

From Corollary \ref{cor1:wbound}, we observe that for
$\tau^{\prime}\leq s$
\begin{equation}
W_i(\tau+\tau^{\prime})\geq W-1,\label{eq4:pfclmkey}
\end{equation}
since using \eqref{eq3:pfclmkey} it is easy to check that
$$\tau^{\prime}~\leq~ s ~\leq~ g^{(-1)}(W/10) ~\leq~
\frac{g^{(-1)}\left(\expo{\log^2 W}\right)}{2W}\qquad\mbox{for large
enough $W$.}$$

For the bound of $g(A^j_i(\tau+\tau^{\prime}))$, we obtain that for
$\tau^{\prime}\leq s$
\begin{equation}
g(A^j_i(\tau+\tau^{\prime}))\leq W/5,\label{eq5:pfclmkey}
\end{equation}
using 1-Lipschitz property of $A^j_i(\cdot)$ and
$$\tau^{\prime}~\leq~ s~\stackrel{(a)}{\leq}~ g^{(-1)}(W/10)~\stackrel{(b)}{\leq}~ g^{(-1)}(W/5)-g^{(-1)}(W/10),$$
where (a) is from \eqref{eq3:pfclmkey} and (b) is due to
$g^{(-1)}(x)\geq g^{(-1)}(x/2)\cdot 2$ for large enough $x$.

\paraG{Second step: Martingale construction} This part of
the proof is similar to that stated in Section \ref{sec:pflem2}.
We consider a modified network Markov chain
where all the Markovian rules are same as the original chain except
for $\bW(\tau^{\prime})=\bW(\tau^{\prime}-1)$ for $\tau^{\prime} >
\tau+s$ i.e.\ $\bW(\cdot)$ is fixed after time $\tau+s$.
This modification does not affect the distribution of
$g(A^j_i(\tau+s))$; it is merely for guaranteeing \eqref{eq4:pfclmkey}
all $\tau\geq 0$.

Now define random time $T_0=\tau< T_1<T_2<\dots$ such that $T_m$ is
the $m$th time when $A^j_i(\cdot)$ is updated since time $\tau$
i.e.\ $B^j_i(T_m-1)\geq 2$ and $B^j_i(T_m)=0$. Further, define for
$m\geq0$,
$$Y_m:=\begin{cases}
\expo{g^{(-1)}(W)-A^j_i(T_{m})}& \mbox{if}~~ T_{m-1}\leq \tau+s~~ \mbox{or}~~m=0 \\
\alpha\cdot Y_{m-1}& \mbox{otherwise}
\end{cases},
$$where $\alpha = \frac{e}{4} + \frac{3}{4e} \in (0,1)$.
We shall establish that for all $m\geq 1$,
\begin{equation}\label{eq6:pfclmkey}
\E\,[Y_{m+1}~|~ \mathcal{F}_m]\leq \alpha\cdot Y_m,
\end{equation}
where $\mathcal{F}_m$ denotes the filtration containing $Y_k,T_k$
for $0\leq k\leq m$. Note that \eqref{eq6:pfclmkey} is trivial if
$T_{m}> \tau+ s$ by definition of $Y_m$. When $T_{m}\leq
\tau+s$, we observe that
\begin{align*}
&\E\,[Y_{m+1}~|~ \mathcal{F}_m]\\
&\quad=\E\left[\expo{g^{(-1)}(W)-A^j_i(T_{m+1})}~|~ \mathcal{F}_m\right]\\
&\quad\stackrel{(a)}{\leq}\frac34\cdot \expo{g^{(-1)}(W)-A^j_i(T_{m})-1}+\frac14\cdot \expo{g^{(-1)}(W)-A^j_i(T_{m})+1}\\
&\quad=\alpha \cdot \expo{g^{(-1)}(W)-A^j_i(T_{m})}\\
&\quad=\alpha \cdot Y_m,
\end{align*}
where for (a) we use
\begin{eqnarray*}
&&\Pr[A^j_i(T_{m+1})=A^j_i(T_{m})+1]~=~\Pr[B^j_i(T_m-1)\geq g(A^j_i(T_{m-1}))]\\
&&\qquad\stackrel{(b)}{\geq}\Pr[B^j_i(T_m-1)\geq W/5]
~=~1-\Pr[B^j_i(T_m-1)< W/5]\\
&&\qquad\geq 1-\sum_{k=1}^{W/5}\Pr[B^j_i(T_m-1)= k]\\
&&\qquad\geq 1-\sum_{k=1}^{W/5}\Pr[\mbox{$i$ stops to attempt at time $T_m-1$}]\\
&&\qquad\stackrel{(c)}{\geq} 1-\sum_{k=1}^{W/5}\frac1{W-1}~\geq~
\frac34,
\end{eqnarray*}
where (b) and (c) are from \eqref{eq5:pfclmkey} and
\eqref{eq4:pfclmkey}, respectively.

From \eqref{eq6:pfclmkey}, $\{Z_m:=Y_m/\alpha^{m-1},m\geq1\}$
becomes a sub-martingale with respect to $\mathcal{F}_m$. If we
define a stopping time $m^*$ as $m^*=\inf\{m:T_m> \tau+s\}$,
$$\E[Z_{m^*}]\stackrel{(a)}{\leq} \E[Z_{1}]=\E[Y_1]\stackrel{(b)}{\leq}
Y_0\cdot e= \expo{g^{(-1)}(W)-A^j_i(\tau)+1},$$ where (a) and (b) is
from the Doob's optional stopping theorem and 1-Lipschitz property
of $A^j_i(\cdot)$. Using the above inequality and Markov's
inequality, we have
\begin{eqnarray}
&&\frac{Y_{m^*}}{\alpha^{m^*-1}}~=~Z_{m^*}~\leq~
\expo{g^{(-1)}(W)-A^j_i(\tau)+1}\cdot
\log \bQ_{\max},\notag\\
&&\qquad\mbox{with probability at least}~1-\frac1{\log
\bQ_{\max}}=1-o(1).\label{eq8:pfclmkey}
\end{eqnarray}
Finally, it follows that
\begin{align*}
A^j_i(\tau+s)
~&=~A^j_i(m^*-1)~\geq~ A^j_i(m^*)-1\\
&=~g^{(-1)}(W)-(g^{(-1)}(W)-A^j_i(T_{m^*}))-1\\
&=~g^{(-1)}(W)-\log Y_{m^*}-1\\
&\stackrel{(a)}{\geq}~ g^{(-1)}(W)- \Big(g^{(-1)}(W)-A^j_i(\tau)+1+\log\log \bQ_{\max}-(m^*-1)\log\frac1{\alpha}\Big)-1\\
&\stackrel{(b)}{\geq}~A^j_i(\tau)+\frac{s}{\log^{n+2} \bQ_{\max}}\log\frac1{\alpha}-\log\log \bQ_{\max}-2\\
&\stackrel{(c)}{\geq}~g^{(-1)}(W/20),
\end{align*}
where (c) is due to the choice of $s=g^{(-1)}(W/20)\cdot
\log^{n+3} \bQ_{\max}$ and large enough $\bQ_{\max}$; (a) and (b)
hold with probability $1-o(1)$ from \eqref{eq8:pfclmkey} and the
Proposition \ref{clm:mstarchernoff}, respectively. This completes the proof of Lemma \ref{clm:key}.
\end{proof}
\begin{proposition}\label{clm:mstarchernoff}
$$\Pr\left[m^*-1 \geq \frac{s}{\log^{n+2} \bQ_{\max}}\right]~= ~1-o(1).$$
\end{proposition}
\begin{proof}
We start by defining random variable $U_{\tau^{\prime}}$.
$$U_{\tau^{\prime}}=
\begin{cases}
1&\mbox{if $A^j_i(\cdot)$ is updated at time $\tau^{\prime}$}\\
0&\mbox{otherwise}
\end{cases},\quad\qquad\mbox{for}~\tau^{\prime}\in[\tau+1,\tau+s].$$
In other words, $U_{\tau^{\prime}}=1$ only if
$B^j_i(\tau^{\prime}-1)\geq 2$ and $B^j_i(\tau^{\prime})=0$. By
definition of $U_{\tau^{\prime}}$ and $m^*$,
$$m^*-1=\sum_{\tau^{\prime}=\tau+1}^{\tau+s} U_{\tau^{\prime}}.$$
Since $W_{\max}(\tau^{\prime})\leq\cW_{\max}=O(\log \bQ_{\max})$ for
$\tau^{\prime}\leq\tau+s$ (cf.\ \eqref{eq:wminmax}), the same
arguments in the proof of Proposition \ref{clm5} leads to the
following bound for the expectation of $m^*-1$.
$$\E[m^*-1]~=~\E\left[\sum_{\tau^{\prime}=\tau+1}^{\tau+s} U_{\tau^{\prime}}\right]~=~
\Omega\left(\frac{s}{(\cW_{\max})^{n+1}}\right)~=~
\Omega\left(\frac{s}{\log^{n+1} \bQ_{\max}}\right).$$

Now we define random variable $Z_{\tau^{\prime}}$ as
$$Z_{\tau^{\prime}}=\E\left[\sum_{\tau^{\prime}=\tau+1}^{\tau+s} U_{\tau^{\prime}}~\Big|~
U_{\tau+1},\dots,U_{\tau^{\prime}-1}  \right],$$ where
$\tau^{\prime}\in[\tau+1, \tau+s+1]$. Hence, it is easy to
observe that
$$Z_{\tau+1}=\E[m^*-1]\qquad\mbox{and}\qquad Z_{\tau+s+1}=m^*-1.$$
Further, $\{Z_{\tau^{\prime}}:\tau^{\prime}\in[\tau+1, \tau+s+1]\}$
forms a martingale with bounded increment i.e.\
$|Z_{\tau^{\prime}}-Z_{\tau^{\prime}+1}|\leq 1$. Therefore, the
statement of Proposition \ref{clm:mstarchernoff} follows by applying
the Azuma's inequality to the martingale $\{Z_{\tau^{\prime}}\}$.
\end{proof}

\end{document}